\documentclass[11pt]{article}

\usepackage[preprint]{acl}

\usepackage[T1]{fontenc}
\usepackage[utf8]{inputenc}
\usepackage{tgtermes}
\usepackage{microtype}
\usepackage[english]{babel}
\usepackage{CJKutf8}
\newcommand{\jp}[1]{\begin{CJK*}{UTF8}{min}#1\end{CJK*}}

\usepackage{amsmath}
\usepackage{amssymb}
\usepackage{graphicx}
\usepackage{booktabs}
\usepackage{tabularx}
\usepackage{array}
\usepackage{multirow}
\usepackage{makecell}
\usepackage[table]{xcolor}
\usepackage{listings}
\usepackage{float}
\usepackage{dblfloatfix}
\usepackage{placeins}
\usepackage{enumitem}
\usepackage{cuted}

\lstdefinestyle{qwenprompt}{
    basicstyle=\ttfamily\scriptsize,
    frame=single,
    rulecolor=\color{black!45},
    backgroundcolor=\color{black!2},
    breaklines=true,
    breakatwhitespace=false,
    columns=fullflexible,
    keepspaces=true,
    showstringspaces=false,
    linewidth=\columnwidth,
    xleftmargin=0pt,
    xrightmargin=0pt,
    framexleftmargin=3pt,
    framexrightmargin=3pt,
    framesep=3pt,
    aboveskip=3pt,
    belowskip=5pt
}

\newcolumntype{C}{>{\centering\arraybackslash}c}
\newcolumntype{D}{>{\columncolor{gray!6}\centering\arraybackslash}c}

\newcolumntype{L}{>{\raggedright\arraybackslash}p{2.45cm}}
\newcolumntype{Y}{>{\centering\arraybackslash}X}
\newcolumntype{G}{>{\columncolor{gray!10}\centering\arraybackslash}X}

\usepackage{fancyvrb}

\title{DialInstruct-TTS: Interactive Expressive Speech Generation via \\ Multi-Turn Dialogue with Fine-Grained Temporal Control}

\author{
    \textbf{Wangzixi Zhou\textsuperscript{1}},
    \textbf{Bagus Tris Atmaja\textsuperscript{1}}
    \textbf{Sakriani Sakti\textsuperscript{1}},
    \\
    \textsuperscript{1}Nara Institute of Science and Technology, Japan \\
    \texttt{zhou.wangzixi.az3@naist.ac.jp}, \\
    \texttt{ssakti@is.naist.jp}, \texttt{bagus.tris@naist.ac.jp}
}

\begin{document}
\maketitle
\begin{abstract}
Existing instruction-based TTS systems primarily support one-shot, utterance-level control, where a single instruction determines the speaking style of an entire utterance. As a result, even minor refinements require rewriting the instruction and regenerating the full utterance, limiting fine-grained interactive control. We propose DialInstruct-TTS, an interactive multi-turn expressive TTS framework that enables users to iteratively refine generated speech through natural-language dialogue until the desired style is achieved. At each dialogue turn, a Qwen-based planner integrates the dialogue history and latest user instruction into a complete TemporalPlan, while a temporal acoustic control module provides fine-grained control over emotion, speaking rate, volume, and phonation. Reference-latent revision preserves non-target regions across turns, allowing users to modify only the requested attributes or temporal spans while retaining previously satisfactory speech. Subjective evaluations demonstrate improved global instruction following, accurate local style realization, and effective non-target preservation. Compared with conventional one-shot interaction requiring full instruction restatement and utterance regeneration, DialInstruct-TTS substantially improves perceived convenience and interaction comfort. Audio samples are available at \url{https://anonymous.4open.science/w/eacl-demo-17C5/}.
\end{abstract}

\section{Introduction}

Recent text-to-speech (TTS) research has moved beyond predefined
emotion or style labels
~\cite{Cho_2025,11385230,liu2025uddettsunifyingdiscretedimensional,tang23_interspeech, zhou2026onesizefitsallpersonalizedculturallyadaptive}
toward natural-language control of expressive speech. Natural-language
instructions can describe emotion, speaking rate, volume, phonation,
and other speaking attributes more flexibly. However, most existing
instruction-based TTS systems
~\cite{10534832,du2025cosyvoice3inthewildspeech,hu2026qwen3ttstechnicalreport}
remain limited to one-shot, utterance-level control: when the result is
unsatisfactory, users must restate the instruction and regenerate the
entire utterance.

This interaction is poorly suited to fine-grained refinement. Even when
only a short span or a single attribute requires modification,
full-utterance regeneration may alter regions that were already
satisfactory. Real-world voice production is instead iterative: after
listening to a previous take, a director may ask that ``this part''
sound angrier, that ``the same part'' be weakened, or that one region
be changed while the rest is retained ~\citep{10.1145/3706598.3714263}. An
interactive TTS should also interpret such feedback in
context, revise only the requested region, and preserve previously accepted content.

\begin{figure}[t]
    \centering
    \includegraphics[width=0.8\columnwidth]{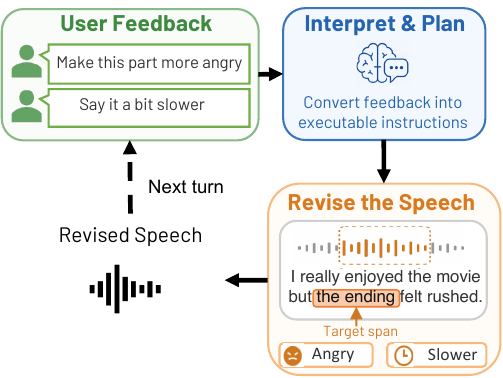}
    \caption{Overview of DialInstruct-TTS}
    \label{fig:overview}
\end{figure}

\begin{figure*}[t]
    \centering
    \includegraphics[width=\textwidth]{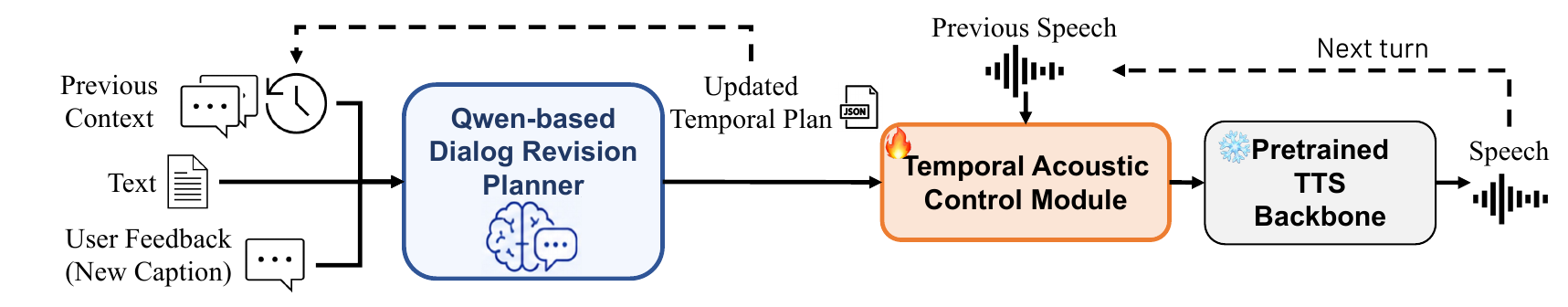}
    \caption{Architecture of DialInstruct-TTS}
    \label{fig:pipeline}
\end{figure*}

We propose DialInstruct-TTS, an interactive multi-turn expressive TTS framework that
allows users to progressively refine generated speech through
natural-language dialogue. At each turn, a Qwen-based Dialogue
Revision Planner ~\cite{qwen3technicalreport} integrates the dialogue
history, previous control state, and latest instruction into a resolved
global caption, a delta plan, and an updated complete TemporalPlan.
This enables context-dependent and relative instructions without
requiring users to restate the complete control specification.
A Temporal Acoustic Control Module realizes the TemporalPlan over a
largely frozen flow-matching TTS backbone, providing span-level control
over emotion, speaking rate, volume, and phonation. Frame-local
cross-attention and attribute-specific controllers apply each change to
its designated temporal region while maintaining the global speaking
style.

To preserve satisfactory regions across dialogue turns, we further introduce
reference-latent revision. The target and transition regions are updated
according to the current instruction, whereas non-target regions retain
their previous acoustic realization. Subjective evaluations demonstrate
improved global instruction following, accurate local style control,
and effective preservation of non-target regions. Compared with
one-shot interaction, DialInstruct-TTS also substantially
improves perceived convenience and interaction comfort.

Our contributions are threefold:
\begin{itemize}
    \item We propose DialInstruct-TTS, an interactive multi-turn
    expressive TTS framework for progressive refinement through
    context-dependent feedback without full instruction restatement.

    \item We develop a Qwen-based Dialogue Revision Planner and a
    Temporal Acoustic Control Module for state tracking, fine-grained
    span-level revision, and non-target preservation.

    \item Experiments and user studies show that DialInstruct-TTS
    effectively controls expressive speech and is strongly preferred
    over one-shot instruction for iterative refinement.
\end{itemize}


\section{Related Work}

\subsection{Instruction-Based Expressive TTS}

Instruction-based expressive TTS uses natural-language descriptions
to control emotion, speaking rate, volume, and other vocal attributes,
providing a more flexible interface than predefined style labels
~\cite{leng2024prompttts, 10448173,10.1145/3664647.3681680,liu23t_interspeech,ji-etal-2025-controlspeech}. Most existing studies have primarily focused on
English, while Japanese-specific instruction-controllable TTS systems
remain limited. Irodori-TTS is a representative Japanese-oriented
expressive TTS system ~\cite{irodori-tts}. Multilingual systems such as
Qwen3-TTS and CosyVoice3 also support Japanese synthesis together with
natural-language style control ~\cite{hu2026qwen3ttstechnicalreport,du2025cosyvoice3inthewildspeech}.

However, these systems generally apply style instructions to the
entire utterance and interpret each request as an independent
generation. To the best of our knowledge, existing instruction-based
TTS systems do not explicitly support span-level temporal control that
modifies only a specified region while preserving previously accepted
non-target regions.

\subsection{Speech Editing and Iterative Revision}

Recent speech-editing methods can modify selected regions of an
existing utterance while preserving the surrounding speech. For
example, VoiceCraft performs local editing through speech-token
infilling, but primarily focuses on changing \textit{what is spoken}
~\cite{peng-etal-2024-voicecraft}. Text-guided voice editing instead modifies acoustic
attributes while keeping the linguistic content unchanged. VoxEditor,
for instance, performs relative voice-attribute editing from textual
prompts, but is mainly formulated as a one-shot, utterance-level
editing task ~\cite{10950106}.
Iterative speech refinement has only recently been explored.
Multi-interaction TTS accepts successive textual directions to
progressively refine the speaking style, but primarily manipulates an
utterance-level style representation and identifies position-specific
control within a sentence as a remaining challenge
~\cite{kanagawa25_ssw}.


\section{DialInstruct-TTS}
\label{sec:method}

The proposed DialInstruct-TTS is a multi-turn temporal reference-revision framework that
modifies only the requested text spans and acoustic attributes while
preserving non-target regions. It consists of a Dialogue Revision
Planner for instruction interpretation and state management, and a
Temporal Acoustic Control Module for localized acoustic revision.

\subsection{Qwen3-Based Dialogue Revision Planner}
\label{sec:dialog_planner}

We locally deploy Qwen3-8B ~\cite{qwen3technicalreport} as the Dialogue
Revision Planner without updating its parameters. A structured system
prompt, provided in Appendix~\ref{sec:qwen_prompt}, instructs the model
to resolve context-dependent feedback, identify revision operations, and
generate executable control plans. The prompt defines the supported
controls, revision rules, and output schema, and is combined with
few-shot examples, JSON-constrained generation, and output validation.
At interaction turn \(k\), the planner receives the source text, the
current user feedback \(u^{(k)}\), and the complete planner output from
the previous turn. The previous output is set to \texttt{null} in the
first turn. 

The previous output contains the resolved global caption, the most recent
delta plan, and all currently active local controls. Using this state,
Qwen3-8B resolves contextual expressions such as ``the same part'' and
``the previously modified region,'' and identifies revision operations
including adding, strengthening, weakening, retaining, replacing, and
removing a control. It then jointly generates the updated caption
\(c^{(k)}\), the current delta plan
\(\Delta\mathcal{P}^{(k)}\), and the complete plan
\(\mathcal{P}^{(k)}\). The previous planner output therefore serves as
explicit revision memory, allowing users to modify earlier results
without restating the target span or previously accepted controls.
The structured output format is detailed in
Appendix~\ref{sec:structured_planner_output}.

\subsection{Temporal Acoustic Control Module}
\label{sec:temporal_acoustic_module}
\begin{figure}[h]
    \centering
    \includegraphics[width=\columnwidth]{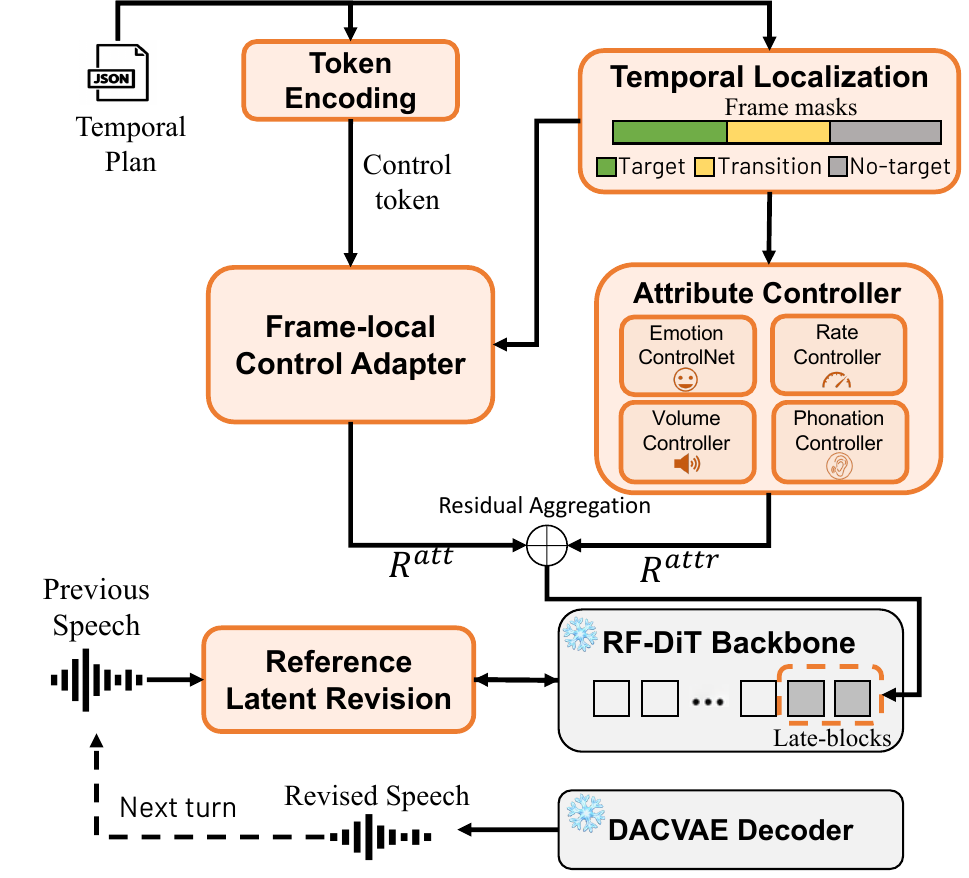}
    \caption{Architecture of the Temporal Acoustic Control Module}
    \label{fig:acoustic}
\end{figure}

We adopt Irodori-TTS ~\cite{irodori-tts} as the backbone and augment it
with the proposed Temporal Acoustic Control Module. As shown in
Figure~\ref{fig:acoustic}, the module comprises TemporalPlan encoding
and localization, frame-local attribute control, and reference-latent
revision.

The RF-DiT backbone ~\cite{10377858} and DACVAE decoder
~\cite{polyak2025moviegencastmedia,semantic-dacvae-japanese-32dim} remain frozen, and
only the newly introduced modules are optimized following a
parameter-efficient adaptation strategy ~\cite{pmlr-v97-houlsby19a}.
Flow matching has been widely applied to speech synthesis ~\cite{10448291, NEURIPS2023_2d8911db,chen-etal-2025-f5}. The detailed training objectives and loss configuration are provided in
Appendix~\ref{sec:training_setting}.

\subsubsection{Token Encoding and Temporal Localization}
\label{sec:temporal_localization}

The Token Encoding module converts each TemporalPlan item
\(\bar{p}_i=(c_i,v_i,s_i,\tau_i,[a_i,b_i),\boldsymbol{\xi}_i)\)
into a control token:
\(k_i=E_c(c_i)+E_v(v_i)+E_{\tau}(\tau_i)+W_s s_i\),
\noindent where \(c_i\), \(v_i\), \(s_i\), and \(\tau_i\) denote the control
type, value, strength, and application type, respectively;
\([a_i,b_i)\) is the target text range, and
\(\boldsymbol{\xi}_i\) contains optional continuous parameters.
Here, \(E_c\), \(E_v\), and \(E_{\tau}\) are learned embeddings for the
control type, value, and application type, and \(W_s\) projects the
control strength into the token space. The resulting sequence
\(Z_P=[k_1,\ldots,k_N]\) represents what should be controlled.

The Temporal Localization module maps each target span to the acoustic
latent timeline using MFA alignment during training. The evaluated
revision implementation also uses MFA-derived locations at inference;
its window diagnostics are described in Appendix~\ref{sec:objective_localization}. It constructs target,
transition, and non-target masks, \(M^{\mathrm{loc}}=M^{\mathrm{tar}}\lor M^{\mathrm{tr}}\) \noindent where \(M^{\mathrm{tar}}\) marks the controlled frames,
\(M^{\mathrm{tr}}\) covers the boundary context, and
\(M^{\mathrm{non}}\) marks the regions to be preserved.

For each acoustic frame \(t\), we further construct a frame-wise
condition:

\begin{equation}
\begin{aligned}
b_t={}&
W_xH_{\pi(t)}^{\mathrm{text}}
+\phi_{\mathrm{tok}}(r_t^{\mathrm{tok}})
+\phi_{\mathrm{span}}(r_t^{\mathrm{span}}) \\
&+E_{\mathrm{rate}}(q_t)
+\phi_s(\log\rho_t),
\end{aligned}
\label{eq:frame_token_condition}
\end{equation}

\noindent where \(b_t\) is the conditioning vector for acoustic frame \(t\).
The function \(\pi(t)\) identifies its aligned text token,
\(H_{\pi(t)}^{\mathrm{text}}\) is the corresponding text representation,
and \(W_x\) projects it into the acoustic hidden space.
The learned encoders \(\phi_{\mathrm{tok}}\) and
\(\phi_{\mathrm{span}}\) represent the relative position of frame \(t\)
within its aligned token and target span, respectively.
The remaining terms encode the speaking-rate category \(q_t\) and the
logarithmic duration ratio \(\rho_t\).

\subsubsection{Frame-Local Control and Attribute-Specific Realization}
\label{sec:frame_local_realization}

Localized control is applied to the later RF-DiT blocks through two
complementary paths. The Control Cross-Attention Adapter interprets
the TemporalPlan semantics and binds them to the target frames, while
the Attribute Controller performs the corresponding emotion,
speaking-rate, volume, and phonation modifications. The residuals from
the two paths are subsequently aggregated and injected into the frozen
RF-DiT backbone.

\paragraph{Control Cross-Attention Adapter}
For each acoustic frame $t$, its hidden state $h_t$ serves as the
query, while the TemporalPlan control tokens serve as keys and values.
A frame--control mask $M_{t,i}$ restricts each frame to the control
items active at its position.

The masked cross-attention output is combined with control-specific
low-rank residuals~\cite{hu2022lora} to produce the control-semantic residual

\begin{equation}
R_t^{\mathrm{att}}
=
g_{\mathrm{CA}}\alpha_t
\left(
W_o a_t
+
\sum_i M_{t,i}r_{t,i}
\right),
\label{eq:attention_residual}
\end{equation}

\noindent where $a_t$ is the masked cross-attention output, $r_{t,i}$ is the
low-rank residual associated with control item $i$, $\alpha_t$ denotes
the control strength, and $g_{\mathrm{CA}}$ is the adapter gate.

\paragraph{Attribute Controller}
The Attribute Controller consists of independent branches with
separate acoustic realization parameters, reducing interference among
emotion, duration, energy, and phonation controls.

\begin{itemize}[
    leftmargin=*,
    noitemsep,
    topsep=2pt
]

    \item \textbf{Emotion}
    The Temporal Emotion ControlNet generates a smooth global-to-local-to-global frame-wise trajectory conditioned on emotion classes, 
    intensity, and span geometry. Following ControlNet ~\cite{10377881}, a
    trainable branch initialized from the final four frozen RF-DiT blocks
    injects emotion residuals into the corresponding backbone blocks.

    \item \textbf{Speaking rate}
    The Rate Controller conditions both local duration allocation in the
    duration predictor and acoustic generation in RF-DiT. At inference,
    it expands or compresses the target span and remaps the frame masks,
    positional features, and reference latent according to the revised
    latent length.

    \item \textbf{Volume}
    The Volume Controller converts the requested gain in decibels into a
    normalized continuous condition and scales a masked acoustic residual
    over the target region. This preserves the direction and magnitude of
    amplification or attenuation while limiting energy changes outside the
    specified span.
    
    \item \textbf{Phonation}    
    The Phonation Controller uses a dedicated whisper branch conditioned on
    the target mask, current and noisy hidden states, and the flow timestep.
    It produces flow-step-dependent phonation residuals, modeling
    whisper-specific voice-quality changes separately from volume instead
    of approximating whispering as simple energy attenuation.
\end{itemize}

The attribute-specific residuals and the final hidden-state update are

\begin{equation}
\begin{aligned}
R_t^{\mathrm{attr}}
&=
\sum_{j\in\mathcal{A}}
g_jR_t^{(j)}, \\
H_t'
&=
H_t
+
R_t^{\mathrm{att}}
+
R_t^{\mathrm{attr}},
\end{aligned}
\label{eq:residual_aggregation}
\end{equation}

\noindent where $\mathcal{A}$ contains the emotion, rate, volume, and phonation
branches, and $g_j$ is the corresponding controller gate. All
residuals are restricted by their frame-level masks and injected only
into the last third of the RF-DiT backbone. 

\subsubsection{Reference-Latent Revision}

Using the previous speech only as a global speaker reference is
insufficient to preserve non-target regions across multiple turns. We
therefore formulate revision as masked flow inpainting and explicitly
clamp the non-target latent trajectory at every sampling step.

Let $r$ denote the previous-turn reference latent aligned to the
current timeline, and let $M^{\mathrm{loc}}$ denote the union of the
target and transition regions. The RF-DiT backbone first produces a
freely updated latent $z_{\tau'}^{\mathrm{free}}$, after which masked
recomposition is performed as

\begin{equation}
\begin{aligned}
z_{\tau'}
={}&
M^{\mathrm{loc}}
\odot
z_{\tau'}^{\mathrm{free}}
\\
&+
\left(
1-M^{\mathrm{loc}}
\right)
\odot
\left[
(1-\tau')r+\tau'\epsilon
\right].
\end{aligned}
\label{eq:reference_clamping}
\end{equation}

The target and transition regions follow the new instruction, while the
non-target regions are projected onto the previous-turn same-noise
trajectory after every flow step. This explicitly preserves the accepted
speech. The resulting latent is passed directly to the next interaction
turn, thereby avoiding repeated conversions between latent
representations and waveforms.

\section{Experiments}

\subsection{Experimental Settings}

\paragraph{Data construction}
To construct Japanese training data for temporal local control, we use
neutral Japanese speech as the base and create
neutral-to-local-style-to-neutral utterances. This is necessary because
recordings that differ only within a local acoustic region are scarce.
We consider eight target styles: \textit{anger}, \textit{fear},
\textit{happy}, \textit{sad}, \textit{fast}, \textit{slow},
\textit{loud}, and \textit{whisper}.
For \textit{fast}, \textit{slow}, and \textit{loud}, MFA
~\cite{mcauliffe17_interspeech} locates the target span, and
SoX\footnote{\url{https://sourceforge.net/projects/sox/}} applies
time-scale or gain modification only to the corresponding PCM segment.
For \textit{whisper}, aligned whisper segments from JVS
~\cite{takamichi2019jvscorpusfreejapanese} are inserted into neutral utterances. For the
four emotion categories, JVNV recordings ~\cite{10418504} provide
emotional references, while Irodori-TTS generates the corresponding
neutral utterances. The aligned emotional spans are then inserted into
the neutral speech. Waveform replacement constructs the training targets. The historical
rate outputs in the additional objective comparison use a separate
MFA-guided PCM editing path, rather than the neural rate controller
(Appendix~\ref{sec:objective_stitch}).
We construct 1k utterances per category,  8k utterances in total, each approximately 8--10 seconds long. Captions are
automatically generated from speaker attributes and target styles. For
example, a female \textit{anger} sample is represented as
\jp{女性の声で、怒った口調で話してください}
(\textit{josei no koe de, okotta kuchō de hanashite kudasai}; ``Please speak in an angry tone using a female voice.'').
All audio is resampled to 48~kHz.

\paragraph{Baselines}
For global instruction-following comparisons, we use the official implementations and released
checkpoints of two state-of-the-art large-scale multilingual TTS systems:
Qwen3-TTS
~\cite{hu2026qwen3ttstechnicalreport}\footnote{\url{https://huggingface.co/Qwen/Qwen3-TTS-12Hz-1.7B-CustomVoice}}
and CosyVoice3
~\cite{du2025cosyvoice3inthewildspeech}\footnote{\url{https://github.com/FunAudioLLM/CosyVoice}}.
We additionally include Irodori-TTS
~\cite{irodori-tts}\footnote{\url{https://github.com/Aratako/Irodori-TTS}},
which serves as the backbone of our DialInstruct-TTS. All baseline systems
are fine-tuned on the same training data for the same number of
optimization steps.

Qwen3-TTS and CosyVoice3 support natural-language instruction control
and are pretrained on large-scale multilingual speech corpora that
include Japanese data. Irodori-TTS is an instruction-controllable
Japanese TTS system. This setup therefore compares our method with
general-purpose instruction-following systems. Since these systems
lack matched local-editing mechanisms, we additionally compare with an
Irodori-based cut-and-stitch editor (Appendix~\ref{sec:objective_stitch}).

\paragraph{Training Setup}
We train 174M temporal-control parameters for around 20k steps on one
NVIDIA RTX PRO 6000, yielding a 791M acoustic model. With the frozen
8B Qwen3 planner, the full system contains 9B parameters.
Further details are provided in Appendix~\ref{sec:training_setting}.

\subsection{Evaluation}

All listening tests involved 30 native Japanese listeners. Participant
and interface details are provided in
Appendix~\ref{sec:interface_human}. We used pairwise preference and
five-point rating tests following established practices
~\cite{wells24_interspeech,naderi20_interspeech}.

\subsubsection{Global Style Evaluation}

We evaluate global instruction following under eight conditions:
four emotions (\textit{happy}, \textit{fear}, \textit{sad}, and
\textit{anger}), two speaking rates (\textit{fast} and \textit{slow}),
and two phonation or volume styles (\textit{loud} and
\textit{whisper}). Each condition is tested with one female and one
male voice, yielding 16 cases per model and 64 samples across four
systems. All systems use the same text, speaker condition, and Japanese
instruction.

Listeners rate each sample on two 5-point scales:

\begin{itemize}[
    leftmargin=*,
    noitemsep,
    topsep=2pt
]
    \item \textbf{MOS:} Naturalness of the generated speech.
    \item \textbf{Match:} Agreement between the generated speaking style
    and the given instruction.
\end{itemize}

We additionally compute Japanese character
error rate (CER) using whisper~\cite{pmlr-v202-radford23a} for all systems.

\begin{table*}[!t]
    \centering
    \caption{Results for global style control. \textbf{MOS} measures
    naturalness, and \textbf{Match} measures agreement with the instruction.
    The best result in each row for each metric is shown in \textbf{bold}.}
    \label{tab:global_category_results}
    \footnotesize
    \setlength{\tabcolsep}{4pt}
    \renewcommand{\arraystretch}{1.08}

    \begin{tabular*}{\textwidth}{
        @{\extracolsep{\fill}}
        lcccccccc
        @{}
    }
        \toprule
        &
        \multicolumn{2}{c}{CosyVoice3}
        & \multicolumn{2}{c}{Irodori-TTS}
        & \multicolumn{2}{c}{Qwen3-TTS}
        & \multicolumn{2}{c}{\textbf{DialInstruct-TTS}} \\
        \cmidrule(lr){2-3}
        \cmidrule(lr){4-5}
        \cmidrule(lr){6-7}
        \cmidrule(lr){8-9}

        \textbf{Category}
        & \textbf{MOS $\uparrow$}
        & \textbf{Match $\uparrow$}
        & \textbf{MOS $\uparrow$}
        & \textbf{Match $\uparrow$}
        & \textbf{MOS $\uparrow$}
        & \textbf{Match $\uparrow$}
        & \textbf{MOS $\uparrow$}
        & \textbf{Match $\uparrow$} \\
        \midrule

        Anger
        & 1.604 & 2.958
        & \textbf{4.042} & 3.708
        & 2.896 & 3.292
        & 3.875 & \textbf{4.417} \\

        Fear
        & 3.709 & 2.509
        & \textbf{3.891} & 1.382
        & 3.564 & 3.345
        & 3.836 & \textbf{4.527} \\

        Happy
        & 2.755 & 3.094
        & \textbf{4.038} & 3.453
        & 3.321 & 2.623
        & 3.906 & \textbf{4.604} \\

        Sad
        & 2.038 & 2.981
        & 3.755 & 2.377
        & 3.472 & \textbf{3.981}
        & \textbf{3.830} & 3.925 \\

        \midrule

        Fast
        & 3.755 & \textbf{4.509}
        & 3.642 & 2.283
        & 3.132 & 3.415
        & \textbf{4.208} & 4.302 \\

        Slow
        & 2.234 & 2.277
        & \textbf{3.915} & 3.745
        & 3.809 & 3.213
        & 3.511 & \textbf{4.192} \\

        \midrule

        Loud
        & 2.148 & 2.389
        & \textbf{4.185} & \textbf{4.074}
        & 2.074 & 2.278
        & 4.019 & 3.759 \\

        Whisper
        & 3.098 & 2.510
        & 4.020 & 1.490
        & 3.275 & 2.490
        & \textbf{4.216} & \textbf{4.549} \\

        \midrule

        \textbf{Overall}
        & 2.688 & 2.911
        & \textbf{3.935} & 2.792
        & 3.186 & 3.077
        & 3.930 & \textbf{4.283} \\

        \bottomrule
    \end{tabular*}
\end{table*}

\subsubsection{Temporal Local Control Evaluation}

We evaluate 16 directed global-to-local style contrasts: all 12 ordered
transitions among \textit{happy}, \textit{sad}, \textit{anger}, and
\textit{fear}; \textit{fast} $\leftrightarrow$ \textit{slow}; and
\textit{loud} $\leftrightarrow$ \textit{whisper}. Each contrast is
applied to a target span at the beginning, middle, or end of the
utterance, yielding 48 test conditions.

Because Qwen3-TTS achieves the second-highest overall Match score in the
global style evaluation and performs best among the baseline systems,
we select it as the comparison system. We conduct a blind pairwise
comparison between our DialInstruct-TTS and Qwen3-TTS. Listeners are
shown the instruction, the input text with the target span highlighted,
and two anonymized audio samples. They first select the sample that
better realizes the requested local style while preserving the global
style in the non-target regions. They then rate the selected sample on four 5-point scales:

\begin{itemize}[
    leftmargin=*,
    noitemsep,
    topsep=2pt
]
    \item \textbf{N-MOS:} Overall naturalness.
    \item \textbf{TN-MOS:} Naturalness of the transition between the
    target span and its surrounding regions.
    \item \textbf{TS-MOS:} Accuracy of the requested style within the
    target span.
    \item \textbf{NSP-MOS:} Preservation of the global style in
    non-target regions.
\end{itemize}

Pairwise preference is the primary
system-level metric. The four ratings are conditional on preference
selection and support diagnostic analysis, not unbiased absolute-quality
comparisons between systems.

\subsubsection{Multi-Turn Evaluation}

We compare full restatement and multi-turn revision using the same TTS
model and target states. Full restatement regenerates the utterance
from scratch without the previous speech or latent, whereas multi-turn
revision reuses the previous control state and reference latent.
Full restatement repeats the global style,
target span, and all active local controls at every turn. Multi-turn
revision uses a complete instruction only at the first turn; later turns
specify only newly added or intensified modifications.

We construct four three-turn dialogues using \textit{happy},
\textit{sad}, \textit{anger}, and \textit{fear} as global emotions.
For each participant, two dialogues use audio instructions and two use
text instructions. The modality assignment is randomized while
maintaining a two-to-two balance.

Participants view both revision sequences side by side and rate them on
three 5-point scales:

\begin{itemize}[
    leftmargin=*,
    noitemsep,
    topsep=2pt
]
    \item \textbf{Perceived Convenience:} Ease of expressing revisions
    without repetitive information.
    \item \textbf{Interaction Comfort:} Naturalness and perceived effort
    of the revision process.
    \item \textbf{Modification Retention:} Preservation of previous
    controls while applying new revisions.
\end{itemize}

After all four dialogues, participants select the system they would
prefer for repeated speech revision: System A, System B, or either
system.

\subsubsection{Additional Objective Evaluation}
We supplement the existing listening tests with objective experiments;
no additional human evaluation is conducted. A backbone-matched
Irodori cut-and-stitch editor is compared on 48 historical local-control
outputs, using an independently dev-selected fixed 5~ms crossfade.
Separately, 20 five-turn acoustic revision tasks compare Proposed with a
variant without non-target locking (reference-latent clamping), yielding 200 attempts and 100 task--turn pairs.
These tasks bypass the planner and reuse the previous generated latent.
We measure target/content success and preservation of current non-target,
never-edited, and previously accepted regions. Settings, component
ablations, full objective results, and MFA window diagnostics appear in
Appendix~\ref{sec:objective_evidence}.

\section{Result and Analysis}

\begin{table*}[!t]
    \centering
    \caption{Preference-selected diagnostic ratings for temporal local style control.
    \textbf{Beg.}, \textbf{Mid.}, and \textbf{End} indicate whether the locally controlled span is
    located at the beginning, middle, or end of the utterance.
    \textbf{N-MOS} evaluates the overall naturalness of the generated speech.
    \textbf{TN-MOS} evaluates the naturalness of the acoustic transitions into and
    out of the target span. \textbf{TS-MOS} measures how accurately the requested
    local style is realized within the target span. \textbf{NSP-MOS} measures how
    well the original global style is preserved in the non-target regions. Best scores are shown in \textbf{bold}.}
    \label{tab:temporal_proposed_diagnostic}

    \footnotesize
    \setlength{\tabcolsep}{2pt}
    \renewcommand{\arraystretch}{1.08}

    \begin{tabularx}{\textwidth}{@{}
        L
        !{\color{gray!60}\vrule width 0.35pt}
        YGY
        !{\color{gray!60}\vrule width 0.35pt}
        GYG
        !{\color{gray!60}\vrule width 0.35pt}
        YGY
        !{\color{gray!60}\vrule width 0.35pt}
        GYG
        @{}}
        \toprule

        \multirow{2}{*}{\textbf{Global $\rightarrow$ Local}}
        & \multicolumn{3}{c!{\color{gray!60}\vrule width 0.35pt}}
          {\textbf{N-MOS $\uparrow$}}
        & \multicolumn{3}{c!{\color{gray!60}\vrule width 0.35pt}}
          {\textbf{TN-MOS $\uparrow$}}
        & \multicolumn{3}{c!{\color{gray!60}\vrule width 0.35pt}}
          {\textbf{TS-MOS $\uparrow$}}
        & \multicolumn{3}{c}
          {\textbf{NSP-MOS $\uparrow$}} \\

        \cmidrule(lr){2-4}
        \cmidrule(lr){5-7}
        \cmidrule(lr){8-10}
        \cmidrule(lr){11-13}

        & \textbf{Beg.}
        & \textbf{Mid.}
        & \textbf{End}
        & \textbf{Beg.}
        & \textbf{Mid.}
        & \textbf{End}
        & \textbf{Beg.}
        & \textbf{Mid.}
        & \textbf{End}
        & \textbf{Beg.}
        & \textbf{Mid.}
        & \textbf{End} \\
        \midrule

        happy $\rightarrow$ sad
        & 3.75 & 3.83 & \textbf{4.14}
        & \textbf{3.75} & 3.67 & 3.71
        & 3.50 & 3.83 & \textbf{4.14}
        & 4.12 & \textbf{4.50} & 4.29 \\

        happy $\rightarrow$ fear
        & \textbf{4.17} & 3.83 & 3.88
        & \textbf{4.00} & 3.17 & 3.38
        & \textbf{4.17} & 4.00 & 4.12
        & 3.67 & 4.00 & \textbf{4.75} \\

        happy $\rightarrow$ anger
        & \textbf{3.80} & 3.50 & 3.14
        & \textbf{3.60} & 3.17 & 3.14
        & 4.40 & \textbf{4.50} & 3.00
        & 4.00 & \textbf{4.17} & 3.86 \\

        sad $\rightarrow$ happy
        & \textbf{3.50} & 3.17 & 3.38
        & 2.50 & 2.50 & \textbf{2.75}
        & 4.00 & 3.83 & \textbf{4.38}
        & 3.25 & 4.00 & \textbf{4.25} \\

        sad $\rightarrow$ fear
        & \textbf{3.50} & 3.00 & \textbf{3.50}
        & \textbf{3.17} & \textbf{3.17} & \textbf{3.17}
        & 3.83 & \textbf{4.50} & 4.17
        & \textbf{4.17} & 3.83 & 3.67 \\

        sad $\rightarrow$ anger
        & 3.50 & 3.40 & \textbf{3.86}
        & \textbf{3.33} & 3.20 & 2.71
        & 4.00 & \textbf{4.20} & 4.14
        & 3.33 & \textbf{4.40} & 4.00 \\

        anger $\rightarrow$ happy
        & 3.57 & \textbf{4.40} & 4.33
        & 3.57 & 3.80 & \textbf{4.33}
        & 4.14 & \textbf{4.60} & 4.50
        & 4.14 & 4.20 & \textbf{4.33} \\

        anger $\rightarrow$ sad
        & 3.57 & \textbf{4.00} & \textbf{4.00}
        & 3.57 & 4.00 & \textbf{4.43}
        & 3.86 & 3.86 & \textbf{4.14}
        & 3.14 & \textbf{4.29} & \textbf{4.29} \\

        anger $\rightarrow$ fear
        & 2.83 & 3.00 & \textbf{3.43}
        & \textbf{3.17} & 2.71 & 3.00
        & 3.67 & 4.00 & \textbf{4.14}
        & 3.83 & 4.00 & \textbf{4.14} \\

        fear $\rightarrow$ happy
        & 3.62 & \textbf{4.00} & 3.43
        & 2.75 & \textbf{4.00} & 3.43
        & 3.75 & \textbf{4.00} & 3.71
        & 4.12 & \textbf{4.17} & 3.71 \\

        fear $\rightarrow$ sad
        & 4.17 & \textbf{4.33} & 3.50
        & 4.00 & \textbf{4.33} & 4.17
        & 4.00 & 3.33 & \textbf{4.33}
        & \textbf{4.33} & \textbf{4.33} & 3.50 \\

        fear $\rightarrow$ anger
        & 2.83 & 4.17 & \textbf{4.50}
        & 3.00 & \textbf{3.67} & 3.50
        & 3.83 & 4.33 & \textbf{4.50}
        & 3.50 & \textbf{4.67} & 4.50 \\

        \midrule

        fast $\rightarrow$ slow
        & 4.25 & \textbf{4.60} & 4.29
        & \textbf{5.00} & 4.40 & 4.29
        & \textbf{5.00} & 3.80 & 4.43
        & \textbf{4.75} & 4.40 & 4.71 \\

        slow $\rightarrow$ fast
        & \textbf{4.00} & 3.50 & 3.00
        & 3.33 & 2.67 & \textbf{3.40}
        & 4.22 & 4.17 & \textbf{5.00}
        & 3.67 & 4.17 & \textbf{4.60} \\

        \midrule

        loud $\rightarrow$ whisper
        & \textbf{3.67} & 2.86 & 3.22
        & \textbf{3.50} & 2.43 & 3.11
        & \textbf{4.83} & 4.00 & 3.89
        & \textbf{4.67} & 4.14 & 4.11 \\

        whisper $\rightarrow$ loud
        & 2.71 & 2.33 & \textbf{3.00}
        & \textbf{3.14} & 2.33 & 3.12
        & 3.43 & 3.33 & \textbf{3.50}
        & 4.29 & \textbf{4.50} & 4.25 \\

        \midrule

        \textbf{Overall}
        & 3.58 & 3.57 & \textbf{3.62}
        & 3.40 & 3.27 & \textbf{3.45}
        & 4.00 & 4.03 & \textbf{4.08}
        & 3.90 & \textbf{4.23} & 4.18 \\

        \bottomrule
    \end{tabularx}
\end{table*}

\subsection{Global Style Control Performance}
Table~\ref{tab:global_category_results} shows that DialInstruct-TTS achieves
the highest overall Match score (4.283), with MOS 3.930, close to
Irodori-TTS (3.935). It has the highest Match for \textit{anger},
\textit{fear}, \textit{happy}, \textit{slow}, and \textit{whisper}, while
CosyVoice3 performs best for \textit{fast} (4.509 versus 4.302).
Thus, the gains concern instruction adherence rather than uniform
naturalness superiority. Detailed category results and variances are in
Appendix~\ref{sec:app_specific_global_style}.

DialInstruct-TTS and Irodori-TTS both obtain zero CER on this small
global-style test; Qwen3-TTS and CosyVoice3 obtain 0.0089 and 0.0125,
respectively, mainly from fillers or lexical insertions. We hypothesize
that temporal-control training improves style representations that also
benefit global control, and examine component effects separately below.

\subsection{Temporal Local Control Quality}
In the blind A/B test, DialInstruct-TTS achieves a macro-average
preference of 92.1\%, versus 7.9\% for Qwen3-TTS. This comparison tests
instruction-following TTS, not superiority over specialized editors.
Table~\ref{tab:temporal_proposed_diagnostic} provides conditional ratings
of preferred samples; detailed results are in
Appendices~\ref{sec:app_specific_temporal_proposed}
and~\ref{sec:app_specific_temporal_qwen}.

TS-MOS is 4.00/4.03/4.08 for beginning/middle/end spans, while TN-MOS is
3.40/3.27/3.45. These ratings indicate perceived local-style realization
and transition weaknesses, but do not directly establish frame-boundary
accuracy or unchanged acoustic realization. Speaking-rate control has
the strongest diagnostic ratings; local acceleration introduces more
transition difficulty. Strong emotion contrasts and whisper-to-loud
changes are also challenging, particularly in middle spans with two
boundaries. Synthetic training boundaries are a possible contributor,
rather than an established sole cause.

The matched objective comparison shows a trade-off
(Appendix~\ref{sec:objective_stitch}): on emotion tasks, proposed outputs
have lower CER (1.07\% versus 2.21\%), while stitching has smaller
non-target acoustic distances. On six phonation tasks, whisper accuracy
is 6/6 versus 2/6 and all three non-target distances are lower for the
proposed outputs. The rate comparison uses PCM editing and is not a
neural-controller comparison; no waveform AV evaluator is available for
this comparison.

\subsection{Perceived Usability of Multi-Turn Revision}
\begin{figure}[t]
    \centering
    \includegraphics[width=\linewidth]{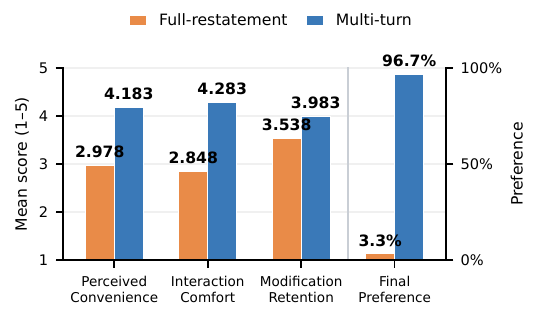}
    \caption{Mean subjective ratings and final preference for
    full-restatement and multi-turn revision. Final preference is
    1/30 versus 29/30, with no ties. Means are shown without error bars;
    numerical results are in Appendix~\ref{sec:app_multiturn_usability}.}
    \label{fig:multiturn_results}
\end{figure}

Figure~\ref{fig:multiturn_results} summarizes the existing user study.
Multi-turn revision improves convenience, comfort, and perceived
modification retention. The corrected convenience score is 4.183;
final preferences are 29/30 for multi-turn revision and 1/30 for full
restatement, with no ties (96.7\% versus 3.3\%). These controlled
three-turn sequences support perceived usability, while acoustic
preservation is evaluated separately below.

Text instructions have slightly higher convenience and comfort ratings
than audio instructions (Appendix~\ref{sec:app_multiturn_usability});
approximately 88\% of participants prefer text feedback. Dialogue here
means context-dependent multi-turn revision and does not require spoken
input. The system relies on user feedback, not autonomous acoustic
self-correction.

\subsection{Component Effects and Cross-Turn Revision}
On the same Irodori backbone (Table~\ref{tab:objective_ablation}),
removing the rate and emotion controllers increases duration error
from 0.06 to 0.91~s and AV distance from 0.11 to 0.64, respectively.
Removing the volume controller increases gain error from 2.7 to
3.1~dB. Without non-target locking (reference-latent clamping), non-target log-F0 and log-energy
RMSE rise from 0.092 to 0.270 and from 0.54 to 0.93.
Removing frame-local attention worsens all five reported metrics:
duration error 0.68~s, AV distance 0.44, non-target log-F0/log-energy
RMSE 0.150/0.66, and gain error 3.1~dB. These component results are
separate from the historical-WAV comparison.

In the five-turn non-target locking ablation, Proposed has higher joint target/content
success than the variant without non-target locking for rate, volume, and rate removal
(Table~\ref{tab:recursive_success}). This joint endpoint combines control
and intelligibility and is not itself a preservation score. Acoustic
preservation is measured against the previous turn, the initial audio in
never-edited regions, and the most recently accepted version of older
edits (Appendix~\ref{sec:objective_recursive}). At turn five, current
non-target mel distance is 4.19~dB for Proposed versus 23.48~dB without
non-target locking ($n=18$ paired sources; difference 19.28~dB, 95\% CI
[17.65, 20.89]). Never-edited regions still drift from the initial
audio (14.75 versus 28.72~dB), and volume success is low. Emotion stages do not use non-target locking in either
variant; the comparison therefore does not establish emotion-only
locking benefits or independently verified emotional success.

\begin{table}[t]
\centering
\small
\setlength{\tabcolsep}{4pt}
\begin{tabular}{lcc}
\toprule
Operation & Proposed & \shortstack{Without non-target\\locking} \\
\midrule
Rate & 15/20 (75\%) & 6/20 (30\%) \\
Volume & 4/20 (20\%) & 1/20 (5\%) \\
Remove rate & 12/20 (60\%) & 5/20 (25\%) \\
\bottomrule
\end{tabular}
\caption{Non-target locking ablation: joint target/content success over five revision turns.
Unavailable assessments remain in the attempt denominator and do not
count as successes.}
\label{tab:recursive_success}
\end{table}

\section{Conclusion}

We proposed DialInstruct-TTS, an interactive multi-turn expressive TTS framework that supports
global style control, temporal local modification, and multi-turn speech
revision. DialInstruct-TTS achieved more consistent global instruction
following than the baselines while maintaining competitive naturalness
and linguistic accuracy. For temporal control, it accurately realized
the requested local style and preserved the surrounding global style,
although transition smoothness remained challenging under strong local
contrasts. Additional matched-editor and recursive experiments reveal
attribute-dependent trade-offs and support reference reuse in the tested
rate/volume stages, while cumulative drift remains. The multi-turn evaluation further showed that incremental instructions
substantially improved perceived convenience and interaction comfort
over full restatement. Participants also generally preferred text-based
instructions to audio-based instructions.










\section*{Limitations}

The current framework has been trained and evaluated only on Japanese
speech, and its cross-lingual generalization remains unexplored.
Emotion control is also limited to four basic categories:
\textit{anger}, \textit{fear}, \textit{happy}, and \textit{sad}.
Future work should extend the framework to additional languages,
compound emotions, and more fine-grained or continuous affective states.
Temporal local control may still reduce overall speech naturalness.
Emotion revision can occasionally introduce speaker drift or abrupt
local boundaries, while speaking-rate control, particularly local
acceleration, may cause rhythmic discontinuities and crowded
articulation. The supported attributes also differ in maturity:
emotion and speaking-rate revision are more stable, whereas volume and
whisper-like phonation remain less reliable. Volume control may cause
clipping or non-target drift, while whispering may sometimes be
perceived as breathiness or reduced energy rather than a distinct
phonation change.
Finally, the current framework assumes fixed linguistic content and
does not support simultaneous lexical editing, arbitrary waveform
editing, or the stable composition of multiple controls within the same
short span. These remain important directions for future work.
The local MOS ratings are selection-conditioned, and the small scripted
user study does not establish open-ended dialogue robustness. The added
acoustic tests bypass the planner, use MFA-derived windows without
independent boundary annotations, and do not validate natural unscripted
recordings. Preservation and boundary-loss ablations remain untested
separately. Automatic acoustic distances do not replace perceptual
naturalness judgments.

\section*{Ethical Considerations}

Qwen3, Qwen3-TTS, and CosyVoice are used under the Apache 2.0 License; Irodori-TTS, Semantic-DACVAE, and MFA under the MIT License; JVNV under CC BY-SA 4.0; JVS under its stated academic and non-commercial research terms; and SoX under the applicable GPL/LGPL terms. All datasets, pretrained models, and software were used in accordance with their respective licenses. The experiments involve general Japanese speech generation and expressive-style control and contain no sensitive, discriminatory, or harmful content.

This research study was reviewed and approved by the Institutional Review Board of [Institution Name Withheld for Double-Blind Review] (approval number and date withheld for anonymous review). Participants were recruited and compensated in accordance with the approved protocol. Participants provided informed consent prior to participation, and no personal information unrelated to the study was collected. Further participant details are provided in Appendix~\ref{sec:interface_human}.

Expressive TTS technology may be misused for voice impersonation or misleading synthetic content. Our experiments were conducted in a controlled research setting, and the released demo is intended solely for academic evaluation. We do not release tools specifically designed for impersonating arbitrary individuals. Although lexical content is fixed, expressive manipulation of harmful source text and accumulated control states remains a potential misuse risk.

AI-assisted tools were used for language support, demo-page design, and assistance organizing experimental reporting and manuscript revisions. The authors are responsible for the research design, experiments, analysis, and verification of manuscript content.




\bibliography{custom}

@ARTICLE{10534832,
  author={Yang, Dongchao and Liu, Songxiang and Huang, Rongjie and Weng, Chao and Meng, Helen},
  journal={IEEE/ACM Transactions on Audio, Speech, and Language Processing}, 
  title={InstructTTS: Modelling Expressive TTS in Discrete Latent Space With Natural Language Style Prompt}, 
  year={2024},
  volume={32},
  number={},
  pages={2913-2925},
  doi={10.1109/TASLP.2024.3402088}}

@misc{hu2026qwen3ttstechnicalreport,
      title={Qwen3-TTS Technical Report}, 
      author={Hangrui Hu and Xinfa Zhu and Ting He and Dake Guo and Bin Zhang and Xiong Wang and Zhifang Guo and Ziyue Jiang and Hongkun Hao and Zishan Guo and Xinyu Zhang and Pei Zhang and Baosong Yang and Jin Xu and Jingren Zhou and Junyang Lin},
      year={2026},
      eprint={2601.15621},
      archivePrefix={arXiv},
      primaryClass={cs.SD},
      url={https://arxiv.org/abs/2601.15621}, 
}

@misc{du2025cosyvoice3inthewildspeech,
      title={CosyVoice 3: Towards In-the-wild Speech Generation via Scaling-up and Post-training}, 
      author={Zhihao Du and Changfeng Gao and Yuxuan Wang and Fan Yu and Tianyu Zhao and Hao Wang and Xiang Lv and Hui Wang and Chongjia Ni and Xian Shi and Keyu An and Guanrou Yang and Yabin Li and Yanni Chen and Zhifu Gao and Qian Chen and Yue Gu and Mengzhe Chen and Yafeng Chen and Shiliang Zhang and Wen Wang and Jieping Ye},
      year={2025},
      eprint={2505.17589},
      archivePrefix={arXiv},
      primaryClass={cs.SD},
      url={https://arxiv.org/abs/2505.17589}, 
}

@inproceedings{10.1145/3706598.3714263,
author = {Brade, Stephen and Anderson, Sam and Kumar, Rithesh and Jin, Zeyu and Truong, Anh},
title = {SpeakEasy: Enhancing Text-to-Speech Interactions for Expressive Content Creation},
year = {2025},
isbn = {9798400713941},
publisher = {Association for Computing Machinery},
address = {New York, NY, USA},
url = {https://doi.org/10.1145/3706598.3714263},
doi = {10.1145/3706598.3714263},
booktitle = {Proceedings of the 2025 CHI Conference on Human Factors in Computing Systems},
articleno = {756},
numpages = {19},
location = {
},
series = {CHI '25}
}

@misc{takamichi2019jvscorpusfreejapanese,
      title={JVS corpus: free Japanese multi-speaker voice corpus}, 
      author={Shinnosuke Takamichi and Kentaro Mitsui and Yuki Saito and Tomoki Koriyama and Naoko Tanji and Hiroshi Saruwatari},
      year={2019},
      eprint={1908.06248},
      archivePrefix={arXiv},
      primaryClass={cs.SD},
      url={https://arxiv.org/abs/1908.06248}, 
}

@ARTICLE{10418504,
  author={Xin, Detai and Jiang, Junfeng and Takamichi, Shinnosuke and Saito, Yuki and Aizawa, Akiko and Saruwatari, Hiroshi},
  journal={IEEE Access}, 
  title={JVNV: A Corpus of Japanese Emotional Speech With Verbal Content and Nonverbal Expressions}, 
  year={2024},
  volume={12},
  number={},
  pages={19752-19764},
  doi={10.1109/ACCESS.2024.3360885}}

@misc{irodori-tts,
  author = {Chihiro Arata},
  title = {Irodori-TTS: A Flow Matching-based Text-to-Speech Model with Emoji-driven Style Control},
  year = {2026},
  publisher = {GitHub},
  journal = {GitHub repository},
  howpublished = {\url{https://github.com/Aratako/Irodori-TTS}}
}

@inproceedings{
    leng2024prompttts,
    title={Prompt{TTS} 2: Describing and Generating Voices with Text Prompt},
    author={Yichong Leng and Zhifang Guo and Kai Shen and Zeqian Ju and Xu Tan and Eric Liu and Yufei Liu and Dongchao Yang and leying zhang and Kaitao Song and Lei He and Xiangyang Li and sheng zhao and Tao Qin and Jiang Bian},
    booktitle={The Twelfth International Conference on Learning Representations},
    year={2024},
    url={https://openreview.net/forum?id=NsCXDyv2Bn}
    }

@INPROCEEDINGS{10448173,
  author={Shimizu, Reo and Yamamoto, Ryuichi and Kawamura, Masaya and Shirahata, Yuma and Doi, Hironori and Komatsu, Tatsuya and Tachibana, Kentaro},
  booktitle={ICASSP 2024 - 2024 IEEE International Conference on Acoustics, Speech and Signal Processing (ICASSP)}, 
  title={PromptTTS++: Controlling Speaker Identity in Prompt-Based Text-To-Speech Using Natural Language Descriptions}, 
  year={2024},
  volume={},
  number={},
  pages={12672-12676},
  doi={10.1109/ICASSP48485.2024.10448173}}

@inproceedings{10.1145/3664647.3681680,
author = {Zhou, Yixuan and Qin, Xiaoyu and Jin, Zeyu and Zhou, Shuoyi and Lei, Shun and Zhou, Songtao and Wu, Zhiyong and Jia, Jia},
title = {VoxInstruct: Expressive Human Instruction-to-Speech Generation with Unified Multilingual Codec Language Modelling},
year = {2024},
isbn = {9798400706868},
publisher = {Association for Computing Machinery},
address = {New York, NY, USA},
url = {https://doi.org/10.1145/3664647.3681680},
doi = {10.1145/3664647.3681680},
booktitle = {Proceedings of the 32nd ACM International Conference on Multimedia},
pages = {554–563},
numpages = {10},
location = {Melbourne VIC, Australia},
series = {MM '24}
}

@inproceedings{peng-etal-2024-voicecraft,
    title = "{V}oice{C}raft: Zero-Shot Speech Editing and Text-to-Speech in the Wild",
    author = "Peng, Puyuan  and
      Huang, Po-Yao  and
      Li, Shang-Wen  and
      Mohamed, Abdelrahman  and
      Harwath, David",
    editor = "Ku, Lun-Wei  and
      Martins, Andre  and
      Srikumar, Vivek",
    booktitle = "Proceedings of the 62nd Annual Meeting of the Association for Computational Linguistics (Volume 1: Long Papers)",
    month = aug,
    year = "2024",
    address = "Bangkok, Thailand",
    publisher = "Association for Computational Linguistics",
    url = "https://aclanthology.org/2024.acl-long.673/",
    doi = "10.18653/v1/2024.acl-long.673",
    pages = "12442--12462"
}

@ARTICLE{10950106,
  author={Sheng, Zheng-Yan and Liu, Li-Juan and Ai, Yang and Pan, Jia and Ling, Zhen-Hua},
  journal={IEEE Transactions on Audio, Speech and Language Processing}, 
  title={Voice Attribute Editing With Text Prompt}, 
  year={2025},
  volume={33},
  number={},
  pages={1641-1652},
  doi={10.1109/TASLPRO.2025.3557193}}

@inproceedings{kanagawa25_ssw,
  title     = {{Multi-interaction TTS toward professional recording reproduction}},
  author    = {Hiroki Kanagawa and Kenichi Fujita and Aya Watanabe and Yusuke Ijima},
  year      = {2025},
  booktitle = {{13th edition of the Speech Synthesis Workshop}},
  pages     = {110--116},
  doi       = {10.21437/SSW.2025-17},
}

@misc{qwen3technicalreport,
      title={Qwen3 Technical Report}, 
      author={Qwen Team},
      year={2025},
      eprint={2505.09388},
      archivePrefix={arXiv},
      primaryClass={cs.CL},
      url={https://arxiv.org/abs/2505.09388}, 
}

@InProceedings{pmlr-v202-radford23a,
  title = 	 {Robust Speech Recognition via Large-Scale Weak Supervision},
  author =       {Radford, Alec and Kim, Jong Wook and Xu, Tao and Brockman, Greg and Mcleavey, Christine and Sutskever, Ilya},
  booktitle = 	 {Proceedings of the 40th International Conference on Machine Learning},
  pages = 	 {28492--28518},
  year = 	 {2023},
  editor = 	 {Krause, Andreas and Brunskill, Emma and Cho, Kyunghyun and Engelhardt, Barbara and Sabato, Sivan and Scarlett, Jonathan},
  volume = 	 {202},
  series = 	 {Proceedings of Machine Learning Research},
  month = 	 {23--29 Jul},
  publisher =    {PMLR},
  url = 	 {https://proceedings.mlr.press/v202/radford23a.html}
}

@inproceedings{mcauliffe17_interspeech,
  title     = {{Montreal Forced Aligner: Trainable Text-Speech Alignment Using Kaldi}},
  author    = {Michael McAuliffe and Michaela Socolof and Sarah Mihuc and Michael Wagner and Morgan Sonderegger},
  year      = {2017},
  booktitle = {{Interspeech 2017}},
  pages     = {498--502},
  doi       = {10.21437/Interspeech.2017-1386},
  issn      = {2958-1796},
}

@article{Cho_2025,
   title={EmoSphere++: Emotion-Controllable Zero-Shot Text-to-Speech Via Emotion-Adaptive Spherical Vector},
   volume={16},
   ISSN={2371-9850},
   url={http://dx.doi.org/10.1109/TAFFC.2025.3561267},
   DOI={10.1109/taffc.2025.3561267},
   number={3},
   journal={IEEE Transactions on Affective Computing},
   publisher={Institute of Electrical and Electronics Engineers (IEEE)},
   author={Cho, Deok-Hyeon and Oh, Hyung-Seok and Kim, Seung-Bin and Lee, Seong-Whan},
   year={2025},
   month=jul, pages={2365–2380} }

@INPROCEEDINGS{11385230,
  author={Zhou, Wangzixi and Atmaja, Bagus Tris and Sakti, Sakriani},
  booktitle={2025 28th Conference of the Oriental COCOSDA International Committee for the Co-ordination and Standardisation of Speech Databases and Assessment Techniques (O-COCOSDA)}, 
  title={Toward Natural Emotional Text-to-Speech System with Fine-Grained Non-Verbal Expression Control}, 
  year={2025},
  volume={},
  number={},
  pages={1-6},
  doi={10.1109/O-COCOSDA68185.2025.11385230}}

@misc{liu2025uddettsunifyingdiscretedimensional,
      title={UDDETTS: Unifying Discrete and Dimensional Emotions for Controllable Emotional Text-to-Speech}, 
      author={Jiaxuan Liu and Yang Xiang and Han Zhao and Xiangang Li and Yingying Gao and Shilei Zhang and Zhenhua Ling},
      year={2025},
      eprint={2505.10599},
      archivePrefix={arXiv},
      primaryClass={cs.LG},
      url={https://arxiv.org/abs/2505.10599}, 
}

@INPROCEEDINGS{10448291,
  author={Mehta, Shivam and Tu, Ruibo and Beskow, Jonas and Székely, Éva and Henter, Gustav Eje},
  booktitle={ICASSP 2024 - 2024 IEEE International Conference on Acoustics, Speech and Signal Processing (ICASSP)}, 
  title={Matcha-TTS: A Fast TTS Architecture with Conditional Flow Matching}, 
  year={2024},
  volume={},
  number={},
  pages={11341-11345},
  doi={10.1109/ICASSP48485.2024.10448291}}

@inproceedings{NEURIPS2023_2d8911db,
 author = {Le, Matthew and Vyas, Apoorv and Shi, Bowen and Karrer, Brian and Sari, Leda and Moritz, Rashel and Williamson, Mary and Manohar, Vimal and Adi, Yossi and Mahadeokar, Jay and Hsu, Wei-Ning},
 booktitle = {Advances in Neural Information Processing Systems},
 doi = {10.52202/075280-0618},
 editor = {A. Oh and T. Naumann and A. Globerson and K. Saenko and M. Hardt and S. Levine},
 pages = {14005--14034},
 publisher = {Curran Associates, Inc.},
 title = {Voicebox: Text-Guided Multilingual Universal Speech Generation at Scale},
 url = {https://proceedings.neurips.cc/paper_files/paper/2023/file/2d8911db9ecedf866015091b28946e15-Paper-Conference.pdf},
 volume = {36},
 year = {2023}
}

@INPROCEEDINGS{10377881,
  author={Zhang, Lvmin and Rao, Anyi and Agrawala, Maneesh},
  booktitle={2023 IEEE/CVF International Conference on Computer Vision (ICCV)}, 
  title={Adding Conditional Control to Text-to-Image Diffusion Models}, 
  year={2023},
  volume={},
  number={},
  pages={3813-3824},
  doi={10.1109/ICCV51070.2023.00355}}

@inproceedings{wells24_interspeech,
  title     = {{Experimental evaluation of MOS, AB and BWS listening test designs}},
  author    = {Dan Wells and Andrea Lorena {Aldana Blanco} and Cassia Valentini and Erica Cooper and Aidan Pine and Junichi Yamagishi and Korin Richmond},
  year      = {2024},
  booktitle = {{Interspeech 2024}},
  pages     = {2695--2699},
  doi       = {10.21437/Interspeech.2024-1778},
  issn      = {2958-1796},
}

@inproceedings{naderi20_interspeech,
  title     = {{An Open Source Implementation of ITU-T Recommendation P.808 with Validation}},
  author    = {Babak Naderi and Ross Cutler},
  year      = {2020},
  booktitle = {{Interspeech 2020}},
  pages     = {2862--2866},
  doi       = {10.21437/Interspeech.2020-2665},
  issn      = {2958-1796},
}

@InProceedings{pmlr-v97-houlsby19a,
  title = 	 {Parameter-Efficient Transfer Learning for {NLP}},
  author =       {Houlsby, Neil and Giurgiu, Andrei and Jastrzebski, Stanislaw and Morrone, Bruna and De Laroussilhe, Quentin and Gesmundo, Andrea and Attariyan, Mona and Gelly, Sylvain},
  booktitle = 	 {Proceedings of the 36th International Conference on Machine Learning},
  pages = 	 {2790--2799},
  year = 	 {2019},
  editor = 	 {Chaudhuri, Kamalika and Salakhutdinov, Ruslan},
  volume = 	 {97},
  series = 	 {Proceedings of Machine Learning Research},
  month = 	 {09--15 Jun},
  publisher =    {PMLR},
  url = 	 {https://proceedings.mlr.press/v97/houlsby19a.html}
}

@inproceedings{liu23t_interspeech,
  title     = {{PromptStyle: Controllable Style Transfer for Text-to-Speech with Natural Language Descriptions}},
  author    = {Guanghou Liu and Yongmao Zhang and Yi Lei and Yunlin Chen and Rui Wang and Lei Xie and Zhifei Li},
  year      = {2023},
  booktitle = {{Interspeech 2023}},
  pages     = {4888--4892},
  doi       = {10.21437/Interspeech.2023-1779},
  issn      = {2958-1796},
}

@inproceedings{ji-etal-2025-controlspeech,
    title = "{C}ontrol{S}peech: Towards Simultaneous and Independent Zero-shot Speaker Cloning and Zero-shot Language Style Control",
    author = "Ji, Shengpeng  and
      Chen, Qian  and
      Wang, Wen  and
      Zuo, Jialong  and
      Fang, Minghui  and
      Jiang, Ziyue  and
      Huang, Hai  and
      Wang, Zehan  and
      Cheng, Xize  and
      Zheng, Siqi  and
      Zhao, Zhou",
    editor = "Che, Wanxiang  and
      Nabende, Joyce  and
      Shutova, Ekaterina  and
      Pilehvar, Mohammad Taher",
    booktitle = "Proceedings of the 63rd Annual Meeting of the Association for Computational Linguistics (Volume 1: Long Papers)",
    month = jul,
    year = "2025",
    address = "Vienna, Austria",
    publisher = "Association for Computational Linguistics",
    url = "https://aclanthology.org/2025.acl-long.346/",
    doi = "10.18653/v1/2025.acl-long.346",
    pages = "6966--6981",
    ISBN = "979-8-89176-251-0"
}

@INPROCEEDINGS{10377858,
  author={Peebles, William and Xie, Saining},
  booktitle={2023 IEEE/CVF International Conference on Computer Vision (ICCV)}, 
  title={Scalable Diffusion Models with Transformers}, 
  year={2023},
  volume={},
  number={},
  pages={4172-4182},
  doi={10.1109/ICCV51070.2023.00387}}

@inproceedings{
hu2022lora,
title={Lo{RA}: Low-Rank Adaptation of Large Language Models},
author={Edward J Hu and yelong shen and Phillip Wallis and Zeyuan Allen-Zhu and Yuanzhi Li and Shean Wang and Lu Wang and Weizhu Chen},
booktitle={International Conference on Learning Representations},
year={2022},
url={https://openreview.net/forum?id=nZeVKeeFYf9}
}

@inproceedings{tang23_interspeech,
  title     = {{EmoMix: Emotion Mixing via Diffusion Models for Emotional Speech Synthesis}},
  author    = {Haobin Tang and Xulong Zhang and Jianzong Wang and Ning Cheng and Jing Xiao},
  year      = {2023},
  booktitle = {{Interspeech 2023}},
  pages     = {12--16},
  doi       = {10.21437/Interspeech.2023-1317},
  issn      = {2958-1796},
}

@misc{polyak2025moviegencastmedia,
      title={Movie Gen: A Cast of Media Foundation Models}, 
      author={Adam Polyak and Amit Zohar and Andrew Brown and Andros Tjandra and Animesh Sinha and Ann Lee and Apoorv Vyas and Bowen Shi and Chih-Yao Ma and Ching-Yao Chuang and David Yan and Dhruv Choudhary and Dingkang Wang and Geet Sethi and Guan Pang and Haoyu Ma and Ishan Misra and Ji Hou and Jialiang Wang and Kiran Jagadeesh and Kunpeng Li and Luxin Zhang and Mannat Singh and Mary Williamson and Matt Le and Matthew Yu and Mitesh Kumar Singh and Peizhao Zhang and Peter Vajda and Quentin Duval and Rohit Girdhar and Roshan Sumbaly and Sai Saketh Rambhatla and Sam Tsai and Samaneh Azadi and Samyak Datta and Sanyuan Chen and Sean Bell and Sharadh Ramaswamy and Shelly Sheynin and Siddharth Bhattacharya and Simran Motwani and Tao Xu and Tianhe Li and Tingbo Hou and Wei-Ning Hsu and Xi Yin and Xiaoliang Dai and Yaniv Taigman and Yaqiao Luo and Yen-Cheng Liu and Yi-Chiao Wu and Yue Zhao and Yuval Kirstain and Zecheng He and Zijian He and Albert Pumarola and Ali Thabet and Artsiom Sanakoyeu and Arun Mallya and Baishan Guo and Boris Araya and Breena Kerr and Carleigh Wood and Ce Liu and Cen Peng and Dimitry Vengertsev and Edgar Schonfeld and Elliot Blanchard and Felix Juefei-Xu and Fraylie Nord and Jeff Liang and John Hoffman and Jonas Kohler and Kaolin Fire and Karthik Sivakumar and Lawrence Chen and Licheng Yu and Luya Gao and Markos Georgopoulos and Rashel Moritz and Sara K. Sampson and Shikai Li and Simone Parmeggiani and Steve Fine and Tara Fowler and Vladan Petrovic and Yuming Du},
      year={2025},
      eprint={2410.13720},
      archivePrefix={arXiv},
      primaryClass={cs.CV},
      url={https://arxiv.org/abs/2410.13720}, 
}

@misc{semantic-dacvae-japanese-32dim,
  author = {Chihiro Arata},
  title = {Semantic-DACVAE-Japanese-32dim: Lightweight Audio VAE for Japanese Speech},
  year = {2026},
  publisher = {Hugging Face},
  journal = {Hugging Face repository},
  howpublished = {\url{https://huggingface.co/Aratako/Semantic-DACVAE-Japanese-32dim}}
}

@inproceedings{chen-etal-2025-f5,
    title = "F5-{TTS}: A Fairytaler that Fakes Fluent and Faithful Speech with Flow Matching",
    author = "Chen, Yushen  and
      Niu, Zhikang  and
      Ma, Ziyang  and
      Deng, Keqi  and
      Wang, Chunhui  and
      JianZhao, JianZhao  and
      Yu, Kai  and
      Chen, Xie",
    editor = "Che, Wanxiang  and
      Nabende, Joyce  and
      Shutova, Ekaterina  and
      Pilehvar, Mohammad Taher",
    booktitle = "Proceedings of the 63rd Annual Meeting of the Association for Computational Linguistics (Volume 1: Long Papers)",
    month = jul,
    year = "2025",
    address = "Vienna, Austria",
    publisher = "Association for Computational Linguistics",
    url = "https://aclanthology.org/2025.acl-long.313/",
    doi = "10.18653/v1/2025.acl-long.313",
    pages = "6255--6271",
    ISBN = "979-8-89176-251-0"
}

@misc{zhou2026onesizefitsallpersonalizedculturallyadaptive,
      title={Beyond One-Size-Fits-All: Personalized and Culturally Adaptive Emotional TTS via Interactive Optimization of Individual Emotion Perception Spaces}, 
      author={Wangzixi Zhou and Bagus Tris Atmaja and Sakriani Sakti},
      year={2026},
      eprint={2608.00998},
      archivePrefix={arXiv},
      primaryClass={eess.AS},
      url={https://arxiv.org/abs/2608.00998}, 
}

\appendix

\section{Qwen Prompt}
\label{sec:qwen_prompt}

The system prompt is divided into four blocks for readability. During
inference, these blocks are concatenated and provided to Qwen3-8B as a
single system prompt.

\begin{lstlisting}[style=qwenprompt]
You are a Dialogue Revision Planner for Japanese expressive
text-to-speech synthesis.

Your task is to convert the current user feedback and the previous
TemporalPlan state into an updated executable TemporalPlan.

You receive the following input fields:

1. text
   The complete Japanese source text. Treat it as immutable.
   Every output span must be an exact substring of text.

2. user_feedback
   The user's current revision request. It may introduce a new speaking
   style or refer to a previous modification using expressions such as
   "the same part", "that part", or "make it stronger".

3. previous_temporal_plan
   The complete JSON output from the previous turn. It contains the
   previously resolved global caption, the most recent delta plan, and
   all currently active local controls. In the first turn, this field is
   null.
\end{lstlisting}

\begin{lstlisting}[style=qwenprompt]
The input follows this JSON format:

{
  "text": "complete Japanese source text",
  "user_feedback": "current revision request",
  "previous_temporal_plan": null
}

For later turns, previous_temporal_plan contains the complete JSON output
from the immediately preceding turn.

Processing rules:

1. If previous_temporal_plan is null, treat the request as the first turn
   and construct a new global caption and complete plan from
   user_feedback.

2. Otherwise, recover the previous global caption and all active local
   controls from previous_temporal_plan.

3. Resolve contextual expressions in user_feedback using the most recent
   delta_plan and complete_plan. Expressions such as "the same part" or
   "the previous part" should normally refer to the most recently
   modified span.

4. Determine whether user_feedback modifies the global style, one or
   more local controls, or both.

5. Identify the requested operation for each affected control:
   add, strengthen, weaken, retain, replace, or remove.

6. Apply only the changes explicitly requested in user_feedback.
   Preserve all unrelated global and local controls from the previous
   turn.
\end{lstlisting}

\begin{lstlisting}[style=qwenprompt]
7. delta_plan must contain only the changes made in the current turn.

8. complete_plan must contain all active local controls after applying
   delta_plan.

9. resolved_caption must describe the complete global speaking style
   after the current revision.

10. Do not infer a new control solely from the semantic content of text.
    All revisions must be derived from user_feedback and its context.

11. Every span must appear exactly in text. Do not paraphrase or invent
    a target span.

12. If a contextual reference cannot be resolved reliably, do not guess.
    Set clarification_required to true and preserve the previous state.

13. Every intensity must be between 0.0 and 1.0. Strengthening increases
    the previous intensity, while weakening decreases it. Do not modify
    the intensity of unrelated controls.

Supported controls:

- emotion: happy, sad, anger, fear
- rate: slow, fast
- volume: up, down
- pause_after: short, medium
- phonation: whisper
\end{lstlisting}

\begin{lstlisting}[style=qwenprompt]
Output valid JSON only. Do not output explanations, Markdown, or any
additional text.

{
  "clarification_required": false,
  "clarification_question": "",
  "resolved_caption": "complete Japanese global-style instruction",
  "delta_plan": [
    {
      "operation": "add|strengthen|weaken|retain|replace|remove",
      "span": "exact substring of text",
      "control": "emotion|rate|volume|pause_after|phonation",
      "value": "supported value",
      "intensity": 0.0
    }
  ],
  "complete_plan": [
    {
      "span": "exact substring of text",
      "control": "emotion|rate|volume|pause_after|phonation",
      "value": "supported value",
      "intensity": 0.0
    }
  ]
}
\end{lstlisting}

\section{Structured Planner Output}
\label{sec:structured_planner_output}

We serialize the planner output as a JSON object with three top-level
fields: \texttt{resolved\_caption}, \texttt{delta\_plan}, and
\texttt{complete\_plan}. A representative output is shown below.

\begin{CJK*}{UTF8}{goth}
\begin{Verbatim}[fontsize=\scriptsize,frame=single]
{
  "resolved_caption":
    "全体を悲しそうに話してください。",
  "delta_plan": [
    {
      "operation": "add",
      "span": "大切なお知らせ",
      "control": "rate",
      "value": "slow",
      "intensity": 0.7
    }
  ],
  "complete_plan": [
    {
      "span": "大切なお知らせ",
      "control": "emotion",
      "value": "happy",
      "intensity": 0.6
    },
    {
      "span": "大切なお知らせ",
      "control": "rate",
      "value": "slow",
      "intensity": 0.7
    }
  ]
}
\end{Verbatim}
\end{CJK*}

\noindent\textbf{Romanization and translation.}
The source text is
\textit{Kyō wa taisetsu na oshirase ga arimasu}
(``We have an important announcement today'').
The feedback is
\textit{Sakki no bubun o, mō sukoshi yukkuri hanashite kudasai}
(``Please speak the previous part a little more slowly'').
The resolved caption is
\textit{Zentai o kanashisō ni hanashite kudasai}
(``Please speak the entire utterance in a sad manner''),
and the target span \textit{taisetsu na oshirase} means
``important announcement.''

The \texttt{delta\_plan} records only the newly added slow-rate control,
whereas the \texttt{complete\_plan} retains the previous happy-emotion
control together with the new rate control. Each plan item specifies an
operation, an exact source-text span, a control type and value, and a
continuous intensity.

\section{Software and Preprocessing Configuration}
\label{sec:soft_conf}
We use Montreal Forced Aligner (MFA) version~3.3.9 with the
\texttt{japanese\_mfa} v3.0 acoustic model for phoneme-level alignment.
SoX version~14.4.2 is used for local time-scale and gain modification.
Japanese CER is computed using
\texttt{kotoba-tech/kotoba-whisper-v2.0}. The same preprocessing and
evaluation configurations are applied to all compared systems.
MFA uses the Japanese MFA dictionary. SoX applies the control-specific
time-scale factors and gain values used in data construction.
Kotoba-Whisper is run in Japanese transcription mode with the default
decoding configuration.

\section{Training Setting}
\label{sec:training_setting}

We adopt a modular stage-wise training strategy. Starting from the
frozen Irodori-TTS backbone, we independently optimize the emotion,
shared control-attention, speaking-rate, volume, and phonation modules
and integrate their checkpoints into the final system. Qwen3-8B remains
frozen and is used only for instruction-to-plan conversion.

The 8,000 constructed utterances are split into 7,200 training and 800 validation samples. Evaluation uses separately prepared test conditions. All acoustic stages use AdamW with
$\beta_1=0.9$, $\beta_2=0.999$, weight decay $0.01$, gradient clipping
at $1.0$, and a warmup--stable--decay learning-rate schedule. Training
uses BF16 computation on a single NVIDIA RTX PRO 6000 GPU, with
effective batch sizes obtained through gradient accumulation. The
complete stage-wise training requires approximately 20k optimization
steps and 20 GPU-hours.

\begin{table}[t]
    \centering
    \caption{Learning rates for stage-wise training.}
    \label{tab:stagewise_training}
    \small
    \setlength{\tabcolsep}{8pt}
    \begin{tabular}{lc}
        \toprule
        \textbf{Stage} & \textbf{Learning Rate} \\
        \midrule
        Emotion       & $2\times10^{-6}$ \\
        Control-CA    & $2\times10^{-6}$ \\
        Speaking rate & $2\times10^{-5}$ \\
        Volume        & $5\times10^{-6}$ \\
        Phonation     & $2\times10^{-5}$ \\
        \bottomrule
    \end{tabular}
\end{table}

Each stage combines a target-weighted flow-matching objective with
duration and non-target preservation losses. Attribute-specific
objectives are additionally applied for emotion, speaking rate, volume,
and phonation. The resulting checkpoints are combined during inference
to form the complete temporal-control model.

The frozen backbone retains its original rectified-flow objective. We
introduce additional region-aware objectives to train the
temporal-control and reference-revision modules.
No automated hyperparameter search was conducted; learning rates were
selected through preliminary runs, and checkpoints were chosen based on
validation loss.

\paragraph{Non-target preservation.}
Let $\hat{z}_0$ denote the predicted clean latent and $r$ the
previous-turn reference latent. We constrain the non-target region to
remain close to the accepted reference realization:

\begin{equation}
\mathcal{L}_{\mathrm{pres}}
=
\frac{
\sum_t
M_t^{\mathrm{non}}
\left\|
\hat{z}_{0,t}-r_t
\right\|_1
}{
\sum_t M_t^{\mathrm{non}}
}.
\label{eq:preservation_loss}
\end{equation}

This objective discourages unintended changes outside the target and
transition regions.

\paragraph{Boundary smoothness.}
To avoid abrupt acoustic changes at the boundaries of a revised span,
we match the first-order latent differences of the prediction and the
local teacher target:

\begin{equation}
\mathcal{L}_{\mathrm{bnd}}
=
\frac{
\sum_t
B_t\,
\operatorname{Huber}
\left(
\Delta\hat{z}_{0,t},
\Delta z_{0,t}^{*}
\right)
}{
\sum_t B_t
},
\label{eq:boundary_loss}
\end{equation}

where $B_t$ selects frames around the
target--transition and transition--non-target boundaries.

\paragraph{Emotion control.}
Emotion supervision is computed over the target span. It combines
emotion-class cross-entropy with Smooth-L1 regression for emotion intensity. Let
$\hat{\mathbf{e}}_{\mathrm{AV}}=[\hat{a},\hat{v}]$ and
$\mathbf{e}_{\mathrm{AV}}^{*}=[a^{*},v^{*}]$. The emotion-control
objective is

\begin{equation}
\begin{aligned}
\mathcal{L}_{\mathrm{emo}}
={}&
\operatorname{CE}
\left(
\hat{y}_{\mathrm{emo}},
y_{\mathrm{emo}}^{*}
\right)
\\
&+
\lambda_{\mathrm{AV}}
\operatorname{SmoothL1}
\left(
\hat{\mathbf{e}}_{\mathrm{AV}},
\mathbf{e}_{\mathrm{AV}}^{*}
\right)
\\
&+
\lambda_{\mathrm{int}}
\operatorname{SmoothL1}
\left(
\hat{I},
I^{*}
\right).
\end{aligned}
\label{eq:emotion_loss}
\end{equation}

We additionally use intensity-ordering and pairwise-ranking objectives
to preserve the relative relation among weak, medium, and strong
emotion conditions. Target-span latent and moment matching are used as
auxiliary regularization.

\paragraph{Speaking-rate control.}
For tokens inside the controlled span, the predicted durations are
matched to the aligned teacher durations in the logarithmic domain:

\begin{equation}
\mathcal{L}_{\mathrm{rate}}
=
\frac{
\sum_n m_n
\left|
\log \hat{d}_n-\log d_n^{*}
\right|
}{
\sum_n m_n
},
\label{eq:rate_loss}
\end{equation}

where $m_n$ selects the target-span tokens.

\paragraph{Volume control.}
Volume supervision matches the target-region energy change relative to
the reference speech:

\begin{equation}
\mathcal{L}_{\mathrm{vol}}
=
\left|
\Delta\log \hat{E}_{\mathrm{tar}}
-
\Delta\log E_{\mathrm{tar}}^{*}
\right|.
\label{eq:volume_loss}
\end{equation}

This objective encourages the requested local gain change while
avoiding unnecessary energy drift in the surrounding regions.

\paragraph{Phonation control.}
For whisper-like phonation, a target-span classifier distinguishes
normal and whisper realizations:

\begin{equation}
\mathcal{L}_{\mathrm{phon}}
=
\operatorname{CE}
\left(
\hat{y}_{\mathrm{phon}},
y_{\mathrm{phon}}^{*}
\right).
\label{eq:phonation_loss}
\end{equation}

Only the attribute-specific objective corresponding to the active
control is enabled for each training sample.

\section{Detailed Results}
\label{sec:re_spec}

\subsection{Specific Global Style Control Performance}
\label{sec:app_specific_global_style}
\begin{table*}[!t]
    \centering
    \caption{Category-wise global-style ratings, reported as mean
    $\pm$ sample variance. Overall pools all responses, and the best
    mean in each row for each metric is shown in bold.}
    \label{tab:app_global_category_results}

    \footnotesize
    \setlength{\tabcolsep}{4pt}
    \renewcommand{\arraystretch}{1.08}

    \resizebox{\textwidth}{!}{%
    \begin{tabular}{@{}
        l
        cc
        @{\hspace{7pt}}
        cc
        @{\hspace{7pt}}
        cc
        @{\hspace{7pt}}
        cc
        @{}}
        \toprule

        \multirow{2}{*}{\textbf{Category}}
        & \multicolumn{2}{c}{\textbf{CosyVoice3}}
        & \multicolumn{2}{c}{\textbf{Irodori-TTS}}
        & \multicolumn{2}{c}{\textbf{Qwen3-TTS}}
        & \multicolumn{2}{c}{\textbf{DialInstruct-TTS}} \\

        \cmidrule(lr){2-3}
        \cmidrule(lr){4-5}
        \cmidrule(lr){6-7}
        \cmidrule(lr){8-9}

        & \textbf{MOS $\uparrow$}
        & \textbf{Match $\uparrow$}
        & \textbf{MOS $\uparrow$}
        & \textbf{Match $\uparrow$}
        & \textbf{MOS $\uparrow$}
        & \textbf{Match $\uparrow$}
        & \textbf{MOS $\uparrow$}
        & \textbf{Match $\uparrow$} \\
        \midrule

        Anger
        & $1.604 \pm 0.712$
        & $2.958 \pm 1.020$
        & $\mathbf{4.042} \pm 1.062$
        & $3.708 \pm 1.020$
        & $2.896 \pm 1.627$
        & $3.292 \pm 1.020$
        & $3.875 \pm 1.431$
        & $\mathbf{4.417} \pm 0.546$ \\

        Fear
        & $3.709 \pm 0.877$
        & $2.509 \pm 0.662$
        & $\mathbf{3.891} \pm 1.173$
        & $1.382 \pm 0.426$
        & $3.564 \pm 1.102$
        & $3.345 \pm 0.823$
        & $3.836 \pm 1.214$
        & $\mathbf{4.527} \pm 1.032$ \\

        Happy
        & $2.755 \pm 1.304$
        & $3.094 \pm 0.895$
        & $\mathbf{4.038} \pm 0.652$
        & $3.453 \pm 0.752$
        & $3.321 \pm 1.222$
        & $2.623 \pm 1.470$
        & $3.906 \pm 0.818$
        & $\mathbf{4.604} \pm 0.321$ \\

        Sad
        & $2.038 \pm 0.999$
        & $2.981 \pm 1.403$
        & $3.755 \pm 0.804$
        & $2.377 \pm 1.240$
        & $3.472 \pm 1.792$
        & $\mathbf{3.981} \pm 0.942$
        & $\mathbf{3.830} \pm 1.105$
        & $3.925 \pm 1.571$ \\

        \midrule

        Fast
        & $3.755 \pm 1.150$
        & $\mathbf{4.509} \pm 0.562$
        & $3.642 \pm 1.427$
        & $2.283 \pm 1.130$
        & $3.132 \pm 1.463$
        & $3.415 \pm 1.363$
        & $\mathbf{4.208} \pm 0.783$
        & $4.302 \pm 0.522$ \\
        
        Slow
        & $2.234 \pm 0.792$
        & $2.277 \pm 0.596$
        & $\mathbf{3.915} \pm 0.862$
        & $3.745 \pm 0.803$
        & $3.808 \pm 0.984$
        & $3.213 \pm 1.476$
        & $3.511 \pm 1.212$
        & $\mathbf{4.192} \pm 0.680$ \\

        \midrule

        Loud
        & $2.148 \pm 1.223$
        & $2.389 \pm 0.770$
        & $\mathbf{4.185} \pm 0.531$
        & $\mathbf{4.074} \pm 0.749$
        & $2.074 \pm 1.051$
        & $2.278 \pm 1.299$
        & $4.019 \pm 0.736$
        & $3.759 \pm 1.016$ \\

        Whisper
        & $3.098 \pm 1.250$
        & $2.510 \pm 1.135$
        & $4.020 \pm 0.780$
        & $1.490 \pm 0.415$
        & $3.275 \pm 1.043$
        & $2.490 \pm 0.815$
        & $\mathbf{4.216} \pm 1.093$
        & $\mathbf{4.549} \pm 0.892$ \\

        \midrule

        \textbf{Overall}
        & $2.688 \pm 1.585$
        & $2.911 \pm 1.321$
        & $\mathbf{3.935} \pm 0.923$
        & $2.792 \pm 1.802$
        & $3.186 \pm 1.512$
        & $3.077 \pm 1.418$
        & $3.930 \pm 1.068$
        & $\mathbf{4.283} \pm 0.901$ \\

        \bottomrule
    \end{tabular}%
    }
\end{table*}

As shown in Table~\ref{tab:app_global_category_results}, the variance
statistics further indicate that our DialInstruct-TTS achieves not only
high instruction-matching accuracy but also comparatively consistent
ratings. Its overall Match variance is 0.901, lower than those of
CosyVoice3 (1.321), Irodori-TTS (1.802), and Qwen3-TTS (1.418),
indicating less disagreement among listeners. At the category level,
our DialInstruct-TTS yields particularly low Match variances for
\textit{happy} (0.321), \textit{fast} (0.522), and \textit{anger}
(0.546), whereas \textit{sad} shows the largest variance (1.571),
suggesting greater listener-dependent variation for this style. Its
overall MOS variance is 1.068, slightly higher than that of Irodori-TTS
(0.923) but lower than those of CosyVoice3 (1.585) and Qwen3-TTS
(1.512), demonstrating relatively stable naturalness despite stronger
style realization. Notably, Irodori-TTS has the lowest overall MOS
variance but the highest Match variance, indicating consistent
perceived naturalness but more variable instruction matching.

\subsection{Specific Temporal Localization Quality}
\label{sec:app_specific_temporal_proposed}
\begin{table*}[t]
    \centering
    \caption{Five-point temporal-control ratings by target-span position
    (mean $\pm$ variance). Best means are shown in bold.}
    \label{tab:app_temporal_proposed_diagnostic}

    \footnotesize
    \setlength{\tabcolsep}{3pt}
    \renewcommand{\arraystretch}{1.08}

    \resizebox{\textwidth}{!}{%
    \begin{tabular}{@{}
        l
        !{\color{gray!45}\vrule width 0.3pt}
        DCD
        !{\color{gray!45}\vrule width 0.3pt}
        CDC
        !{\color{gray!45}\vrule width 0.3pt}
        DCD
        !{\color{gray!45}\vrule width 0.3pt}
        CDC
        @{}}
        \toprule

        \multirow{2}{*}{\textbf{Global $\rightarrow$ Local}}
        & \multicolumn{3}{c!{\color{gray!45}\vrule width 0.3pt}}
          {\textbf{N-MOS}}
        & \multicolumn{3}{c!{\color{gray!45}\vrule width 0.3pt}}
          {\textbf{TN-MOS}}
        & \multicolumn{3}{c!{\color{gray!45}\vrule width 0.3pt}}
          {\textbf{TS-MOS}}
        & \multicolumn{3}{c}
          {\textbf{NSP-MOS}} \\

        \cmidrule(lr){2-4}
        \cmidrule(lr){5-7}
        \cmidrule(lr){8-10}
        \cmidrule(lr){11-13}

        & \textbf{Beg.}
        & \textbf{Mid.}
        & \textbf{End}
        & \textbf{Beg.}
        & \textbf{Mid.}
        & \textbf{End}
        & \textbf{Beg.}
        & \textbf{Mid.}
        & \textbf{End}
        & \textbf{Beg.}
        & \textbf{Mid.}
        & \textbf{End} \\
        \midrule

        happy $\rightarrow$ sad
        & $3.75 \pm 1.1$
        & $3.83 \pm 2.6$
        & $\mathbf{4.14 \pm 0.5}$
        & $\mathbf{3.75 \pm 0.2}$
        & $3.67 \pm 1.5$
        & $3.71 \pm 1.2$
        & $3.50 \pm 1.4$
        & $3.83 \pm 1.8$
        & $\mathbf{4.14 \pm 1.1}$
        & $4.12 \pm 1.0$
        & $\mathbf{4.50 \pm 0.7}$
        & $4.29 \pm 0.9$ \\

        happy $\rightarrow$ fear
        & $\mathbf{4.17 \pm 1.4}$
        & $3.83 \pm 0.6$
        & $3.88 \pm 1.0$
        & $\mathbf{4.00 \pm 0.4}$
        & $3.17 \pm 1.8$
        & $3.38 \pm 1.4$
        & $\mathbf{4.17 \pm 1.4}$
        & $4.00 \pm 1.2$
        & $4.12 \pm 0.7$
        & $3.67 \pm 1.9$
        & $4.00 \pm 1.2$
        & $\mathbf{4.75 \pm 0.2}$ \\

        happy $\rightarrow$ anger
        & $\mathbf{3.80 \pm 0.7}$
        & $3.50 \pm 1.1$
        & $3.14 \pm 0.8$
        & $\mathbf{3.60 \pm 1.3}$
        & $3.17 \pm 1.4$
        & $3.14 \pm 0.5$
        & $4.40 \pm 0.3$
        & $\mathbf{4.50 \pm 0.3}$
        & $3.00 \pm 1.3$
        & $4.00 \pm 0.5$
        & $\mathbf{4.17 \pm 1.0}$
        & $3.86 \pm 1.1$ \\

        sad $\rightarrow$ happy
        & $\mathbf{3.50 \pm 1.1}$
        & $3.17 \pm 1.4$
        & $3.38 \pm 1.4$
        & $2.50 \pm 1.7$
        & $2.50 \pm 1.9$
        & $\mathbf{2.75 \pm 1.1}$
        & $4.00 \pm 1.1$
        & $3.83 \pm 2.2$
        & $\mathbf{4.38 \pm 1.1}$
        & $3.25 \pm 1.4$
        & $4.00 \pm 1.2$
        & $\mathbf{4.25 \pm 0.5}$ \\

        sad $\rightarrow$ fear
        & $\mathbf{3.50 \pm 1.1}$
        & $3.00 \pm 1.6$
        & $\mathbf{3.50 \pm 1.1}$
        & $\mathbf{3.17 \pm 1.4}$
        & $\mathbf{3.17 \pm 0.6}$
        & $\mathbf{3.17 \pm 1.4}$
        & $3.83 \pm 0.6$
        & $\mathbf{4.50 \pm 0.3}$
        & $4.17 \pm 0.6$
        & $\mathbf{4.17 \pm 1.0}$
        & $3.83 \pm 0.6$
        & $3.67 \pm 1.1$ \\

        sad $\rightarrow$ anger
        & $3.50 \pm 1.1$
        & $3.40 \pm 1.3$
        & $\mathbf{3.86 \pm 0.8}$
        & $\mathbf{3.33 \pm 1.5}$
        & $3.20 \pm 1.7$
        & $2.71 \pm 1.2$
        & $4.00 \pm 1.2$
        & $\mathbf{4.20 \pm 0.7}$
        & $4.14 \pm 1.1$
        & $3.33 \pm 0.7$
        & $\mathbf{4.40 \pm 0.3}$
        & $4.00 \pm 1.0$ \\

        anger $\rightarrow$ happy
        & $3.57 \pm 1.0$
        & $\mathbf{4.40 \pm 0.3}$
        & $4.33 \pm 1.5$
        & $3.57 \pm 1.3$
        & $3.80 \pm 1.7$
        & $\mathbf{4.33 \pm 0.7}$
        & $4.14 \pm 1.1$
        & $\mathbf{4.60 \pm 0.3}$
        & $4.50 \pm 0.7$
        & $4.14 \pm 1.1$
        & $4.20 \pm 0.7$
        & $\mathbf{4.33 \pm 0.7}$ \\

        anger $\rightarrow$ sad
        & $3.57 \pm 1.3$
        & $\mathbf{4.00 \pm 0.3}$
        & $\mathbf{4.00 \pm 0.3}$
        & $3.57 \pm 1.0$
        & $4.00 \pm 1.0$
        & $\mathbf{4.43 \pm 0.6}$
        & $3.86 \pm 0.5$
        & $3.86 \pm 1.1$
        & $\mathbf{4.14 \pm 1.1}$
        & $3.14 \pm 0.8$
        & $\mathbf{4.29 \pm 0.2}$
        & $\mathbf{4.29 \pm 0.6}$ \\

        anger $\rightarrow$ fear
        & $2.83 \pm 1.4$
        & $3.00 \pm 1.7$
        & $\mathbf{3.43 \pm 0.6}$
        & $\mathbf{3.17 \pm 1.4}$
        & $2.71 \pm 0.9$
        & $3.00 \pm 1.3$
        & $3.67 \pm 1.1$
        & $4.00 \pm 0.0$
        & $\mathbf{4.14 \pm 1.5}$
        & $3.83 \pm 1.4$
        & $4.00 \pm 0.7$
        & $\mathbf{4.14 \pm 0.5}$ \\

        fear $\rightarrow$ happy
        & $3.62 \pm 0.8$
        & $\mathbf{4.00 \pm 1.6}$
        & $3.43 \pm 1.3$
        & $2.75 \pm 1.4$
        & $\mathbf{4.00 \pm 0.4}$
        & $3.43 \pm 1.6$
        & $3.75 \pm 1.6$
        & $\mathbf{4.00 \pm 1.6}$
        & $3.71 \pm 1.6$
        & $4.12 \pm 0.7$
        & $\mathbf{4.17 \pm 1.8}$
        & $3.71 \pm 0.9$ \\

        fear $\rightarrow$ sad
        & $4.17 \pm 0.6$
        & $\mathbf{4.33 \pm 0.3}$
        & $3.50 \pm 0.7$
        & $4.00 \pm 0.4$
        & $\mathbf{4.33 \pm 0.3}$
        & $4.17 \pm 1.0$
        & $4.00 \pm 0.4$
        & $3.33 \pm 1.3$
        & $\mathbf{4.33 \pm 0.7}$
        & $\mathbf{4.33 \pm 0.3}$
        & $\mathbf{4.33 \pm 0.3}$
        & $3.50 \pm 0.3$ \\

        fear $\rightarrow$ anger
        & $2.83 \pm 1.4$
        & $4.17 \pm 0.6$
        & $\mathbf{4.50 \pm 0.5}$
        & $3.00 \pm 2.0$
        & $\mathbf{3.67 \pm 1.5}$
        & $3.50 \pm 4.5$
        & $3.83 \pm 2.2$
        & $4.33 \pm 0.7$
        & $\mathbf{4.50 \pm 0.5}$
        & $3.50 \pm 1.5$
        & $\mathbf{4.67 \pm 0.3}$
        & $4.50 \pm 0.5$ \\

        \midrule

        fast $\rightarrow$ slow
        & $4.25 \pm 0.9$
        & $\mathbf{4.60 \pm 0.3}$
        & $4.29 \pm 1.2$
        & $\mathbf{5.00 \pm 0.0}$
        & $4.40 \pm 0.8$
        & $4.29 \pm 1.2$
        & $\mathbf{5.00 \pm 0.0}$
        & $3.80 \pm 1.2$
        & $4.43 \pm 0.6$
        & $\mathbf{4.75 \pm 0.2}$
        & $4.40 \pm 0.8$
        & $4.71 \pm 0.2$ \\

        slow $\rightarrow$ fast
        & $\mathbf{4.00 \pm 1.0}$
        & $3.50 \pm 2.7$
        & $3.00 \pm 2.5$
        & $3.33 \pm 1.5$
        & $2.67 \pm 1.5$
        & $\mathbf{3.40 \pm 3.3}$
        & $4.22 \pm 0.4$
        & $4.17 \pm 2.6$
        & $\mathbf{5.00 \pm 0.0}$
        & $3.67 \pm 1.5$
        & $4.17 \pm 2.6$
        & $\mathbf{4.60 \pm 0.3}$ \\

        \midrule

        loud $\rightarrow$ whisper
        & $\mathbf{3.67 \pm 2.3}$
        & $2.86 \pm 2.5$
        & $3.22 \pm 1.7$
        & $\mathbf{3.50 \pm 2.7}$
        & $2.43 \pm 1.3$
        & $3.11 \pm 2.1$
        & $\mathbf{4.83 \pm 0.2}$
        & $4.00 \pm 2.0$
        & $3.89 \pm 1.6$
        & $\mathbf{4.67 \pm 0.3}$
        & $4.14 \pm 2.1$
        & $4.11 \pm 0.9$ \\

        whisper $\rightarrow$ loud
        & $2.71 \pm 1.9$
        & $2.33 \pm 1.1$
        & $\mathbf{3.00 \pm 2.0}$
        & $\mathbf{3.14 \pm 1.5}$
        & $2.33 \pm 1.1$
        & $3.12 \pm 1.3$
        & $3.43 \pm 1.3$
        & $3.33 \pm 3.5$
        & $\mathbf{3.50 \pm 1.4}$
        & $4.29 \pm 0.6$
        & $\mathbf{4.50 \pm 0.7}$
        & $4.25 \pm 1.1$ \\

        \midrule

        \textbf{Overall}
        & $3.58 \pm 1.2$
        & $3.57 \pm 1.5$
        & $\mathbf{3.62 \pm 1.2}$
        & $3.40 \pm 1.3$
        & $3.27 \pm 1.4$
        & $\mathbf{3.45 \pm 1.5}$
        & $4.00 \pm 1.0$
        & $4.03 \pm 1.2$
        & $\mathbf{4.08 \pm 1.1}$
        & $3.90 \pm 1.0$
        & $\mathbf{4.23 \pm 0.9}$
        & $4.18 \pm 0.7$ \\

        \bottomrule
    \end{tabular}%
    }
\end{table*}

As shown in Table~\ref{tab:app_temporal_proposed_diagnostic}, the
variance statistics provide a complementary view of temporal-control
stability. Overall, TS-MOS and NSP-MOS show relatively low variances
across positions (1.0--1.2 and 0.7--1.0, respectively), indicating
comparatively consistent listener judgments of target-style realization
and non-target preservation. In contrast, N-MOS and TN-MOS exhibit
higher variances, particularly for middle-position N-MOS (1.5) and
end-position TN-MOS (1.5), suggesting greater disagreement regarding
overall naturalness and transition smoothness. Several difficult
conditions, such as \textit{fear} $\rightarrow$ \textit{anger} at the
end (TN-MOS variance 4.5), \textit{slow} $\rightarrow$ \textit{fast}
at the end (3.3), and \textit{whisper} $\rightarrow$ \textit{loud} in
the middle (TS-MOS variance 3.5), produce particularly variable
ratings. By contrast, \textit{fast} $\rightarrow$ \textit{slow} shows
consistently low variances, confirming that this control direction is
realized more robustly. These results further indicate that target-style
realization and non-target preservation are relatively stable, whereas
transition naturalness remains the main source of listener disagreement.

\subsection{Specific Temporal Localization Quality of Qwen3-TTS}
\label{sec:app_specific_temporal_qwen}
As shown in Table~\ref{tab:qwen_temporal_diagnostic}, the variance
results should be interpreted together with the low preference rate of
Qwen3-TTS. Its overall N-MOS, TN-MOS, TS-MOS, and NSP-MOS variances are
0.686, 0.885, 0.953, and 0.900, respectively, indicating moderate
variability among the Qwen3-TTS samples selected by listeners.
Relatively stable ratings are observed for several conditions, such as
\textit{sad} $\rightarrow$ \textit{happy} (N-MOS and TS-MOS variances
of 0.200), \textit{fear} $\rightarrow$ \textit{anger} (N-MOS variance
of 0.218), and \textit{slow} $\rightarrow$ \textit{fast} (NSP-MOS
variance of 0.250). In contrast, \textit{whisper} $\rightarrow$
\textit{loud} shows particularly high TS-MOS and NSP-MOS variances of
2.250 and 2.917, while \textit{fear} $\rightarrow$ \textit{happy}
yields variances of 2.333 for both TN-MOS and NSP-MOS. These results
suggest that Qwen3-TTS occasionally produces acceptable localized
control, but its performance is inconsistent across style transitions.
Moreover, because Qwen3-TTS was selected in only 7.9\% of the pairwise
trials, these conditional ratings are based on a small subset of its
outputs and should therefore be regarded as diagnostic rather than
representative of its overall temporal-control performance.

\begin{table*}[!t]
    \centering
    \caption{Qwen3-TTS results for temporal local style control.
    Preference rate is the proportion of comparisons in which Qwen3-TTS
    was selected. Ratings are reported as mean $\pm$ sample variance.
    The highest result in each column is shown in bold.}
    \label{tab:qwen_temporal_diagnostic}

    \footnotesize
    \setlength{\tabcolsep}{5pt}
    \renewcommand{\arraystretch}{1.08}

    \resizebox{\textwidth}{!}{%
    \begin{tabular}{@{}
        l
        c
        @{\hspace{8pt}}
        cccc
        @{}}
        \toprule

        \textbf{Global $\rightarrow$ Local}
        & \shortstack{\textbf{Preference}\\\textbf{Rate (\%)}}
        & \textbf{N-MOS $\uparrow$}
        & \textbf{TN-MOS $\uparrow$}
        & \textbf{TS-MOS $\uparrow$}
        & \textbf{NSP-MOS $\uparrow$} \\
        \midrule

        happy $\rightarrow$ sad
        & 4.8
        & $\mathbf{4.250} \pm 0.250$
        & $4.000 \pm 1.333$
        & $\mathbf{4.250} \pm 0.917$
        & $3.750 \pm 1.583$ \\

        happy $\rightarrow$ fear
        & 6.3
        & $4.000 \pm 1.600$
        & $\mathbf{4.333} \pm 0.267$
        & $3.833 \pm 0.567$
        & $3.833 \pm 0.567$ \\

        happy $\rightarrow$ anger
        & 7.7
        & $4.000 \pm 0.667$
        & $3.857 \pm 1.143$
        & $3.429 \pm 2.286$
        & $3.714 \pm 0.905$ \\

        sad $\rightarrow$ happy
        & 6.1
        & $2.800 \pm 0.200$
        & $3.600 \pm 0.300$
        & $2.800 \pm 0.200$
        & $3.600 \pm 1.300$ \\

        sad $\rightarrow$ fear
        & 8.1
        & $3.818 \pm 0.764$
        & $4.091 \pm 0.691$
        & $3.455 \pm 1.073$
        & $3.727 \pm 0.618$ \\

        sad $\rightarrow$ anger
        & 15.0
        & $4.188 \pm 0.562$
        & $3.938 \pm 0.996$
        & $3.625 \pm 1.317$
        & $4.125 \pm 0.783$ \\

        anger $\rightarrow$ happy
        & 6.3
        & $3.833 \pm 0.567$
        & $4.000 \pm 0.800$
        & $4.167 \pm 0.567$
        & $4.167 \pm 0.567$ \\

        anger $\rightarrow$ sad
        & 9.4
        & $3.556 \pm 1.028$
        & $3.889 \pm 0.861$
        & $3.778 \pm 0.944$
        & $3.222 \pm 0.694$ \\

        anger $\rightarrow$ fear
        & 7.9
        & $3.429 \pm 0.619$
        & $3.714 \pm 0.905$
        & $3.714 \pm 0.571$
        & $3.857 \pm 0.476$ \\

        fear $\rightarrow$ happy
        & 3.7
        & $3.667 \pm 1.333$
        & $3.333 \pm 2.333$
        & $4.000 \pm 0.000$
        & $3.333 \pm 2.333$ \\

        fear $\rightarrow$ sad
        & \textbf{16.7}
        & $3.789 \pm 0.509$
        & $3.842 \pm 0.696$
        & $3.526 \pm 0.596$
        & $3.368 \pm 1.468$ \\

        fear $\rightarrow$ anger
        & 11.5
        & $3.727 \pm 0.218$
        & $3.818 \pm 0.764$
        & $3.364 \pm 0.655$
        & $3.727 \pm 0.618$ \\

        \midrule

        fast $\rightarrow$ slow
        & 6.1
        & $4.000 \pm 0.800$
        & $4.167 \pm 0.567$
        & $3.333 \pm 1.467$
        & $3.500 \pm 0.700$ \\

        slow $\rightarrow$ fast
        & 4.1
        & $4.000 \pm 0.667$
        & $3.500 \pm 0.333$
        & $3.250 \pm 0.917$
        & $\mathbf{4.250} \pm 0.250$ \\

        \midrule

        loud $\rightarrow$ whisper
        & 7.6
        & $3.125 \pm 0.696$
        & $3.000 \pm 1.429$
        & $3.625 \pm 1.125$
        & $3.750 \pm 0.500$ \\

        whisper $\rightarrow$ loud
        & 4.9
        & $4.000 \pm 0.667$
        & $2.500 \pm 1.000$
        & $2.750 \pm 2.250$
        & $3.250 \pm 2.917$ \\

        \midrule

        \textbf{Overall}
        & 7.9
        & $3.778 \pm 0.686$
        & $3.794 \pm 0.885$
        & $3.556 \pm 0.953$
        & $3.698 \pm 0.900$ \\

        \bottomrule
    \end{tabular}%
    }
\end{table*}

\subsection{Multi-Turn Usability Details}
\label{sec:app_multiturn_usability}
Table~\ref{tab:multiturn_corrected} gives the numerical values shown in
Figure~\ref{fig:multiturn_results}. Final preferences are 29/30 for
multi-turn revision and 1/30 for full restatement, without ties.
Table~\ref{tab:instruction_modality} separately reports ratings by
instruction modality; these modality means are not additional systems.

\begin{table}[t]
\centering
\small
\setlength{\tabcolsep}{4pt}
\begin{tabular}{lcc}
\toprule
 & Restatement & Multi-turn \\
\midrule
Convenience & 2.978 & 4.183 \\
Comfort & 2.848 & 4.283 \\
Modification retention & 3.538 & 3.983 \\
Final preference & 1/30 & 29/30 \\
\bottomrule
\end{tabular}
\caption{Existing multi-turn user-study results with corrected
convenience and final preference reporting. No ties occurred.}
\label{tab:multiturn_corrected}
\end{table}

\begin{table}[h]
    \centering
    \caption{Multi-turn ratings by instruction modality.}
    \label{tab:instruction_modality}
    \small
    \setlength{\tabcolsep}{6pt}
    \renewcommand{\arraystretch}{1.10}
    \begin{tabular}{lcc}
        \toprule
        \textbf{Instruction}
        & \shortstack{\textbf{Perceived $\uparrow$}\\\textbf{Convenience }}
        & \shortstack{\textbf{Interaction $\uparrow$}\\\textbf{Comfort}} \\
        \midrule
        Audio & 4.176 & 4.147 \\
        Text  & 4.194 & 4.333 \\
        \bottomrule
    \end{tabular}
\end{table}

\section{Additional Objective Evidence}
\label{sec:objective_evidence}

\subsection{Component Ablations}
\label{sec:objective_components}
Table~\ref{tab:objective_ablation} reports the component ablations on the
same Irodori backbone. Duration error is reported in seconds and target
gain error in dB; AV distance, non-target log-F0 RMSE, and non-target
log-energy RMSE retain their original evaluation scales. A dash denotes
an unreported measurement, not zero error. These results are distinct
from the new matched historical-output and recursive experiments.
Removing frame-local attention worsens all reported control and
preservation measurements. No separate preservation-loss or
boundary-loss ablation is reported.

\begin{table*}[t]
\centering
\small
\setlength{\tabcolsep}{6pt}
\begin{tabular}{lccccc}
\toprule
Variant & Abs. duration error (s)$\downarrow$ & AV distance$\downarrow$ & NT log-F0$\downarrow$ & NT log-energy$\downarrow$ & Gain error (dB)$\downarrow$ \\
\midrule
Proposed & 0.06 & 0.11 & 0.092 & 0.54 & 2.7 \\
$-$ Rate controller & 0.91 & --- & --- & --- & --- \\
$-$ Emotion controller & --- & 0.64 & --- & --- & --- \\
$-$ Volume controller & --- & --- & --- & --- & 3.1 \\
$-$ Non-target locking & --- & --- & 0.270 & 0.93 & --- \\
$-$ Frame-local attention & 0.68 & 0.44 & 0.150 & 0.66 & 3.1 \\
\bottomrule
\end{tabular}
\caption{Component ablations. NT denotes non-target acoustic error.
Unreported entries are shown as dashes.}
\label{tab:objective_ablation}
\end{table*}

\subsection{Backbone-Matched Local Regeneration and Stitching}
\label{sec:objective_stitch}
We compare 48 historical proposed WAVs with 48 Irodori-based stitched
outputs; all 96 outputs have valid acoustic measurements. The editor
regenerates the target style, extracts the corresponding span, and
inserts it into the shared source audio. Crossfade is fixed at 5~ms,
selected on independent development data. Table~\ref{tab:objective_stitch}
reports task-level means; CER and accuracy are fractions, and lower
values are better only for error/distance metrics.

The 36 emotion, six rate, and six phonation tasks partition the 48
tasks. The three volume tasks overlap the phonation set and are shown
only as a diagnostic subset. Proposed rate outputs in this comparison
already use MFA-guided direct PCM editing, so their duration comparison
does not establish neural rate-controller superiority. Historical
selected proposed outputs are compared with one preregistered donor
seed per stitch task; there is no equivalent best-of-$N$ sampling claim.

\begin{table*}[t]
\centering
\small
\setlength{\tabcolsep}{6pt}
\begin{tabular}{llrrr}
\toprule
Family & Metric & Paired $n$ & Proposed & Stitch \\
\midrule
Emotion & Full CER$\downarrow$ & 36 & 0.0107 & 0.0221 \\
 & NT log-energy RMSE$\downarrow$ & 36 & 0.7102 & 0.1642 \\
 & NT log-F0 RMSE$\downarrow$ & 36 & 0.0870 & 0.0144 \\
 & NT mel distance (dB)$\downarrow$ & 36 & 9.8119 & 1.8958 \\
 & Target AV distance$\downarrow$ & 0 & --- & --- \\
\midrule
Rate & Full CER$\downarrow$ & 6 & 0.0284 & 0.0284 \\
 & NT log-energy RMSE$\downarrow$ & 6 & 0.1389 & 0.2715 \\
 & NT log-F0 RMSE$\downarrow$ & 6 & 0.0140 & 0.0163 \\
 & NT mel distance (dB)$\downarrow$ & 6 & 1.5157 & 2.3903 \\
 & Target duration error (s)$\downarrow$ & 6 & 0.4983 & 0.2024 \\
\midrule
Phonation & Full CER$\downarrow$ & 6 & 0.0838 & 0.0838 \\
 & NT log-energy RMSE$\downarrow$ & 6 & 0.2309 & 0.3278 \\
 & NT log-F0 RMSE$\downarrow$ & 6 & 0.0315 & 0.0763 \\
 & NT mel distance (dB)$\downarrow$ & 6 & 2.8541 & 3.8570 \\
 & Target whisper accuracy$\uparrow$ & 6 & 1.0000 & 0.3333 \\
\midrule
Volume subset & Full CER$\downarrow$ & 3 & 0.1187 & 0.1187 \\
 & NT log-energy RMSE$\downarrow$ & 3 & 0.1980 & 0.0472 \\
 & NT log-F0 RMSE$\downarrow$ & 3 & 0.0403 & 0.0945 \\
 & NT mel distance (dB)$\downarrow$ & 3 & 2.5878 & 0.6753 \\
 & Target gain error (dB)$\downarrow$ & 3 & 5.9677 & 5.6535 \\
 & Target whisper accuracy$\uparrow$ & 3 & 1.0000 & 0.6667 \\
\bottomrule
\end{tabular}
\caption{Backbone-matched historical-output comparison. Non-target F0
and energy use natural-log scales. Volume is an overlapping subset,
not an independent task family. The gain result also belongs to the
phonation family with the same three paired tasks. No mixed-unit overall
score is computed.}
\label{tab:objective_stitch}
\end{table*}

No output AV evaluator is available for these 48 WAVs, and no emotion
distances are imputed. A lower CER does not establish successful target
emotion. Non-target waveform copying gives stitching a legitimate
preservation advantage; the comparison therefore reports control and
preservation separately. Shared MFA alignment is independent of the
synthesis masks but can align incorrect pronunciations, so CER remains
a separate measurement. Word-phase mapping, frame grids, and pitch
estimation can produce nonzero distances even for unchanged PCM;
exact prefix/suffix checks are retained in the evaluation audit.
Small three- and six-task groups support descriptive findings only.

\subsection{MFA-Derived Window Diagnostics}
\label{sec:objective_localization}
All 48 historical edit windows are exported, including 16 at each
position. The historical locator is Japanese MFA, not a learned span
alignment predictor. The available source/output references use the
same MFA acoustic model, leaving zero independent reference boundaries.
Independent localization MAE, IoU, and failure rate are consequently
unavailable. The 0\% operational export failure rate is not a
localization-accuracy result.

Table~\ref{tab:mfa_window_diagnostics} instead compares selected edit
windows with output-MFA timing. This diagnoses their temporal coverage,
not independent boundary accuracy or target emotion. Neural paths use
native DACVAE frames of 40~ms; the rate path uses PCM sample intervals.
IoU is calculated on a common 25~Hz frame-center grid, clipped to the
declared audio duration. Rate donor-duration replacement and suffix
shifts are recorded on the corresponding output timeline.

\begin{table}[t]
\centering
\small
\setlength{\tabcolsep}{4pt}
\begin{tabular}{lrrrr}
\toprule
Group & $n$ & Start (ms) & End (ms) & IoU \\
\midrule
All & 48 & 89.79 & 147.71 & 0.9486 \\
Beginning & 16 & 26.87 & 113.75 & 0.9616 \\
Middle & 16 & 185.63 & 244.37 & 0.9170 \\
End & 16 & 56.88 & 85.00 & 0.9673 \\
Emotion & 36 & 105.00 & 179.72 & 0.9432 \\
Rate & 6 & 6.67 & 40.00 & 0.9833 \\
Phonation & 6 & 81.67 & 63.33 & 0.9466 \\
\bottomrule
\end{tabular}
\caption{Output-MFA window diagnostics: mean absolute start/end
differences and temporal-coverage IoU. These are not independent
localization accuracy measurements.}
\label{tab:mfa_window_diagnostics}
\end{table}

Emotion masks are saved, whereas historical phonation editable masks
are deterministically reconstructed from recorded inputs/code and
logged editable counts where available. They are not described as
captured runtime tensors. Deterministic replay and current-runtime
checks are distinguished from evidence of the historical execution.

\subsection{Non-Target Locking Ablation over Five Turns}
\label{sec:objective_recursive}
This reference-latent clamping ablation compares Proposed with a variant without non-target locking. Twenty source tasks are revised for five turns with both systems, producing 200 attempts and 100 task--turn pairs. Each sequence
includes rate modification, volume modification, and rate removal in
addition to emotion stages. Planner interpretation is bypassed; each
turn reuses the previous actually generated latent. The revised protocol
adds three distinct spans, per-turn MFA correspondence, absolute-duration
restoration, and a separate volume runtime. These extensions are not
attributed to the earlier local-control evaluation. A precheck reproduces
12 legacy generation latents element-for-element.

Joint success requires rate relative duration error at most 5\% or
volume gain error at most 1~dB, together with full-utterance and target
CER at most 0.20. Rate removal is assessed against the specified
restored duration. Table~\ref{tab:recursive_full_counts} includes all
attempts, without selecting successful outputs. Unavailable joint
assessments remain in the denominator and do not count as successes.

\begin{table*}[t]
\centering
\small
\setlength{\tabcolsep}{5.8pt}
\begin{tabular}{llrrrr}
\toprule
System & Operation & Success/attempts & Confirmed failure & Joint unavailable & Target CER unavailable \\
\midrule
Proposed & Rate & 15/20 & 4 & 1 & 1 \\
 & Volume & 4/20 & 14 & 2 & 2 \\
 & Remove rate & 12/20 & 7 & 1 & 1 \\
\shortstack[l]{Without non-target\\locking} & Rate & 6/20 & 13 & 1 & 1 \\
 & Volume & 1/20 & 17 & 2 & 2 \\
 & Remove rate & 5/20 & 14 & 1 & 3 \\
\bottomrule
\end{tabular}
\caption{Proposed-cohort recursive outcomes. Confirmed failures include
execution failures; without-locking rate removal contains two generation
failures. Target-CER unavailability can overlap failure categories and
must not be added to them. Emotion stages have no independent success
assessment and are not included in these control-success rows.}
\label{tab:recursive_full_counts}
\end{table*}

Preservation has three references: current non-target regions versus
the previous turn, never-edited regions versus the initial audio, and
older accepted edits versus their most recently modified version.
Measurements are non-target mel distance, natural-log F0 RMSE, and
natural-log RMS-energy RMSE. MFA maps corresponding text regions across
turns, so rate-induced suffix shifts are not directly scored as acoustic
drift. Missing regions or common voiced frames are unavailable, not
zero error. Acoustic summaries use available matched task--turn pairs;
each system is not independently filtered to its successful outputs.
For the paired acoustic curves, 95\% intervals use 2,000 cluster
bootstrap resamples of the 20 sources rather than independent turns.

Proposed improves joint success by 45 percentage points for rate, 15 for
volume, and 35 for rate removal. These are descriptive differences;
joint success combines target control and content and does not by
itself establish the size of acoustic preservation gains. Both systems
remain weak on volume, and Proposed still has cumulative drift. Historical
emotion inference has non-target locking disabled in both variants. Removing non-target locking affects the rate/volume stages and rate-preserving re-execution;
these results do not establish emotion-only locking effects.

\begin{table*}[t]
\centering
\small
\setlength{\tabcolsep}{6pt}
\begin{tabular}{llrrrr}
\toprule
Region & Metric$\downarrow$ & Proposed & \shortstack{Without non-target\\locking} & Difference & 95\% CI of difference \\
\midrule
Current non-target & Mel (dB) & 4.19 & 23.48 & 19.28 & [17.65, 20.89] \\
 & Log-F0 (ln Hz) & 0.0257 & 0.3731 & 0.3474 & [0.3013, 0.3912] \\
 & Log-energy (ln RMS) & 0.3491 & 2.0779 & 1.7288 & [1.5345, 1.9201] \\
\midrule
Never edited & Mel (dB) & 14.75 & 28.72 & 13.97 & [12.15, 15.80] \\
 & Log-F0 (ln Hz) & 0.1511 & 0.3903 & 0.2392 & [0.1852, 0.3018] \\
 & Log-energy (ln RMS) & 0.9888 & 1.8279 & 0.8390 & [0.7192, 0.9611] \\
\midrule
Earlier edits & Mel (dB) & 3.55 & 25.19 & 21.64 & [19.95, 23.11] \\
 & Log-F0 (ln Hz) & 0.0170 & 0.3539 & 0.3370 & [0.2884, 0.3921] \\
 & Log-energy (ln RMS) & 0.2701 & 2.1198 & 1.8497 & [1.6113, 2.0897] \\
\bottomrule
\end{tabular}
\caption{Turn-five preservation on 18 jointly metric-valid source pairs,
without filtering by control success. Difference is the variant without non-target locking minus
Proposed; positive values favor Proposed. Intervals use source-cluster bootstrap.
Earlier-edits values average identical valid regions within each pair.
Different units are not combined into an overall score.}
\label{tab:recursive_preservation_turn5}
\end{table*}

\begin{figure*}[t]
\centering
\includegraphics[width=\textwidth]{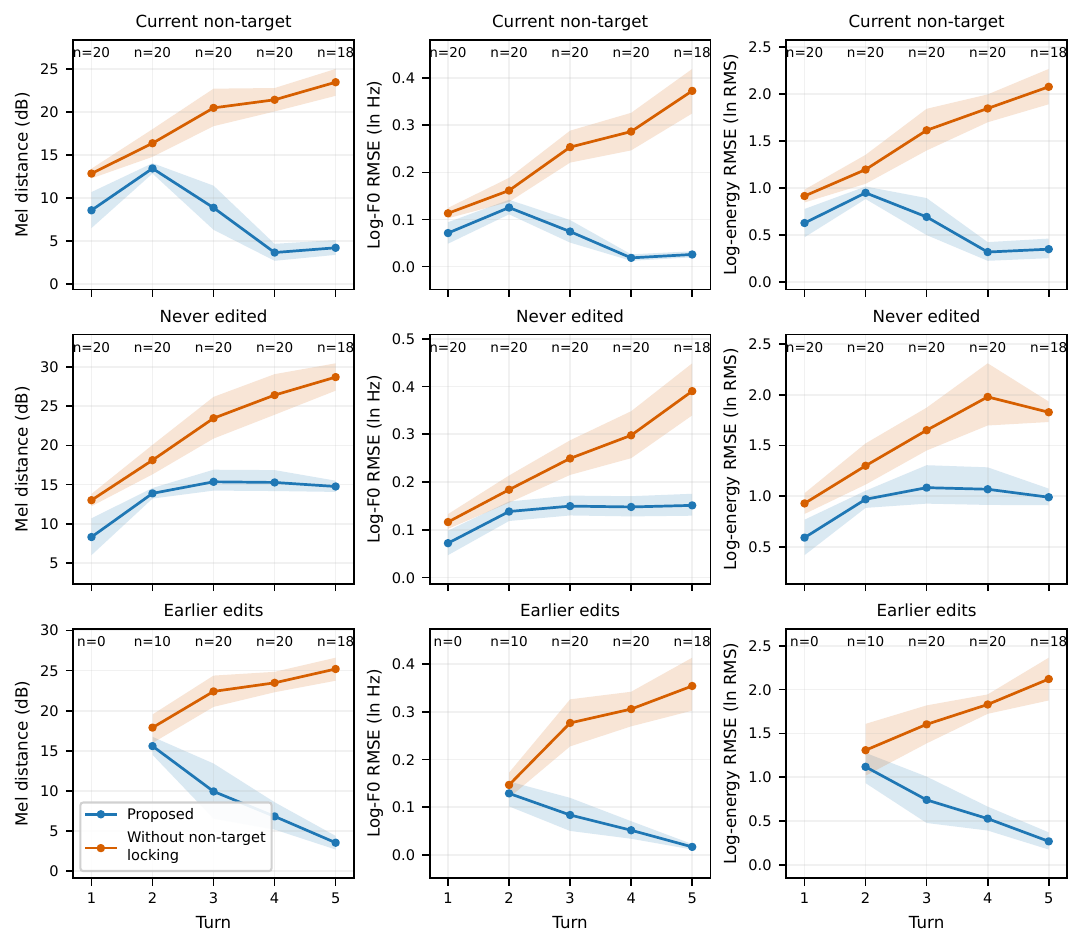}
\caption{Acoustic preservation over five actual revision turns. Curves
use the same metric-valid pairs in both systems; labels show pair counts,
and shading shows 95\% source-cluster bootstrap intervals. Earlier-edits
measurements average corresponding valid regions within a task; turn one
has no earlier edit. Distances are acoustic proxies, not emotion-accuracy
or perceptual-naturalness scores. Emotion generation disables non-target locking
in both variants; differences can reflect accumulated rate/volume edits.}
\label{fig:recursive_preservation}
\end{figure*}

Table~\ref{tab:recursive_preservation_turn5} and
Figure~\ref{fig:recursive_preservation} use all common metric-valid pairs,
not the successful subset. At turn five, all three region types have
smaller mel, log-F0, and log-energy distances for Proposed, with positive
paired-difference intervals. Two without-locking generation failures leave
18 rather than 20 valid pairs; failures remain in the success-rate
denominator. Earlier edits use an equal-region mean within each task,
restricted to identical valid region labels and anchor turns in both
systems. MFA correspondence uses linear phase within aligned words and
excludes pauses. The historical Proposed rate path applies warped-reference
clamping with zero refinement; removing clamping also removes this
reconstruction constraint, while retaining speaker-reference context.
Thus, the contrast evaluates this runtime constraint rather than an
isolated learned neural rate controller.

\paragraph{Successful-subset diagnostic.}
Separately, 31 Proposed records from 16 sources satisfy the joint checks:
15 rate, four volume, and 12 rate-removal records. This outcome-selected
subset has mean full CER 3.14\% and target CER 2.78\%, with current
non-target log-F0 RMSE 0.02436, log-energy RMSE 0.37551, and mel distance
4.30~dB ($n=31$). Never-edited-region mel distance relative to the
initial audio is 12.45~dB ($n=31$); older emotion-region distance relative
to its last edit is 3.59~dB ($n=23$). These conditional measurements
illustrate residual drift after successful edits, not full-cohort
averages or verified emotion retention. Four five-turn sequences have
all tested rate/volume/removal operations passing the joint checks and
all five full-utterance CERs at most 0.20; emotion class/intensity
success remains unassessed.

\section{Human Evaluation}
\label{sec:interface_human}

\begin{figure*}[!t]
    \centering
    \includegraphics[
        width=0.95\textwidth,
        height=0.80\textheight,
        keepaspectratio
    ]{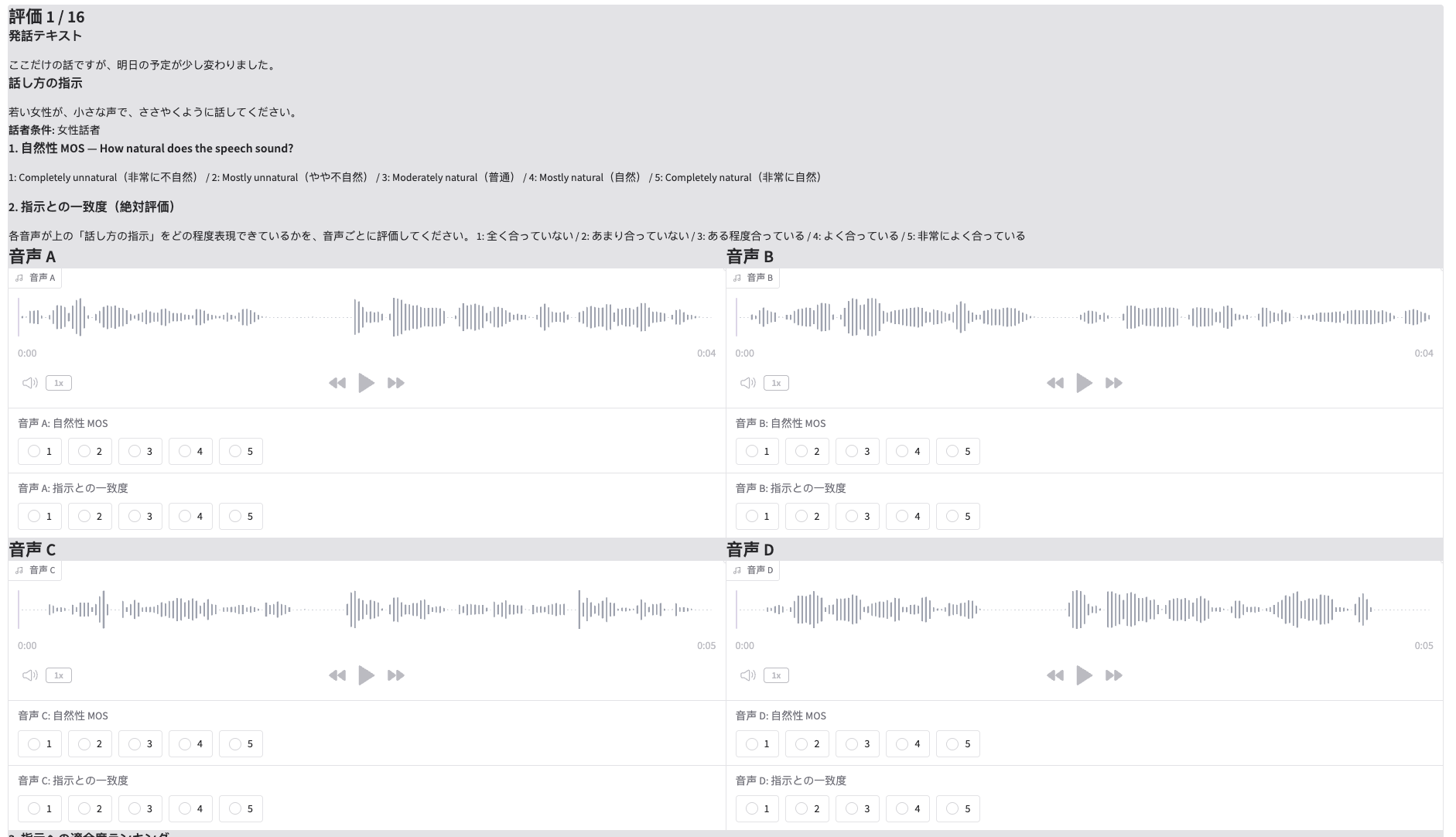}

    \caption{Interface used for the global evaluation.}
    \label{fig:global}
\end{figure*}

\begin{figure*}[!t]

    \centering
    \includegraphics[
        width=0.95\textwidth,
        height=0.80\textheight,
        keepaspectratio
    ]{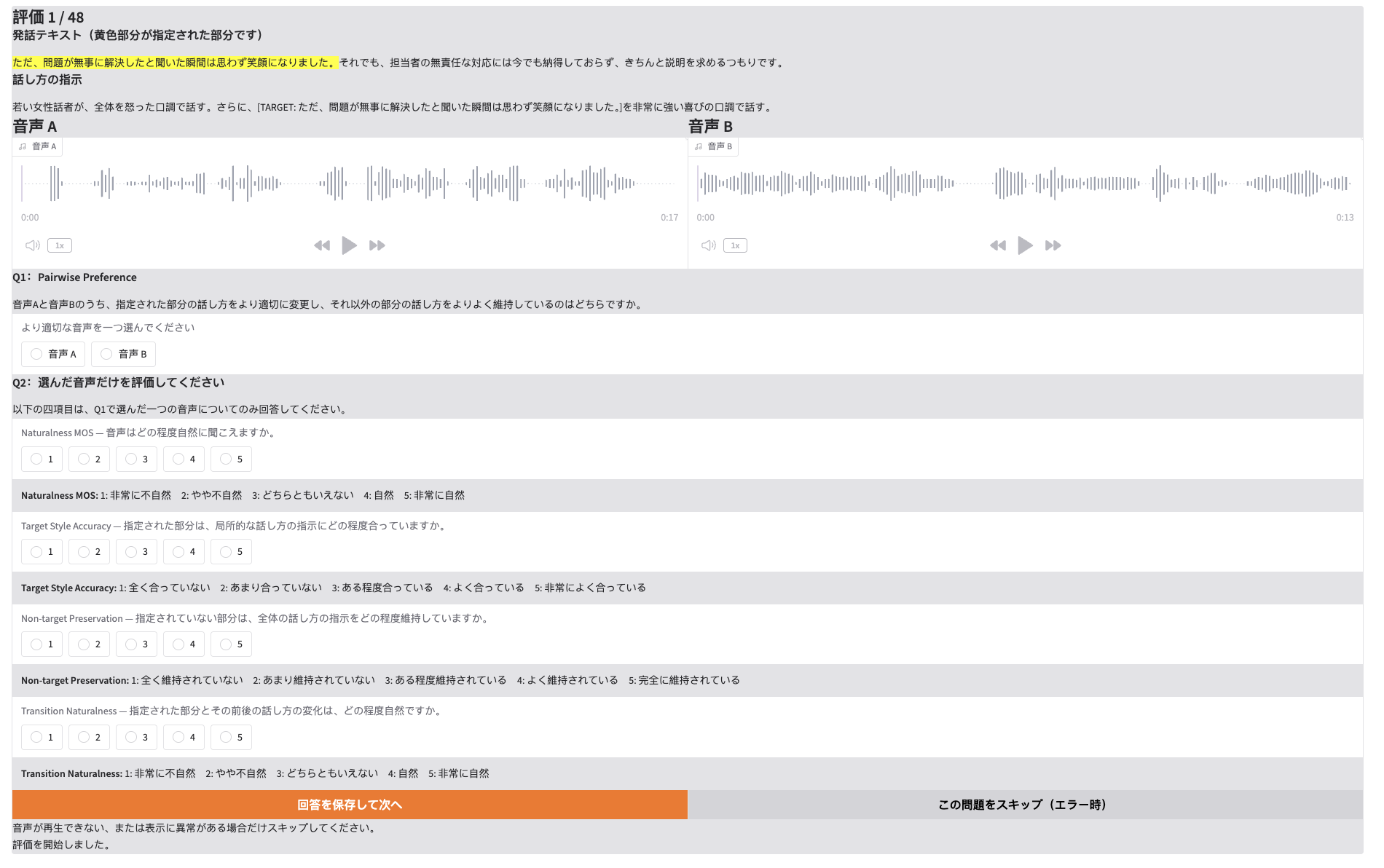}

    \caption{Interface used for the temporal evaluation.}
    \label{fig:tem}
\end{figure*}

\begin{figure*}[!t]
    \centering
    \includegraphics[
        width=0.95\textwidth,
        height=0.80\textheight,
        keepaspectratio
    ]{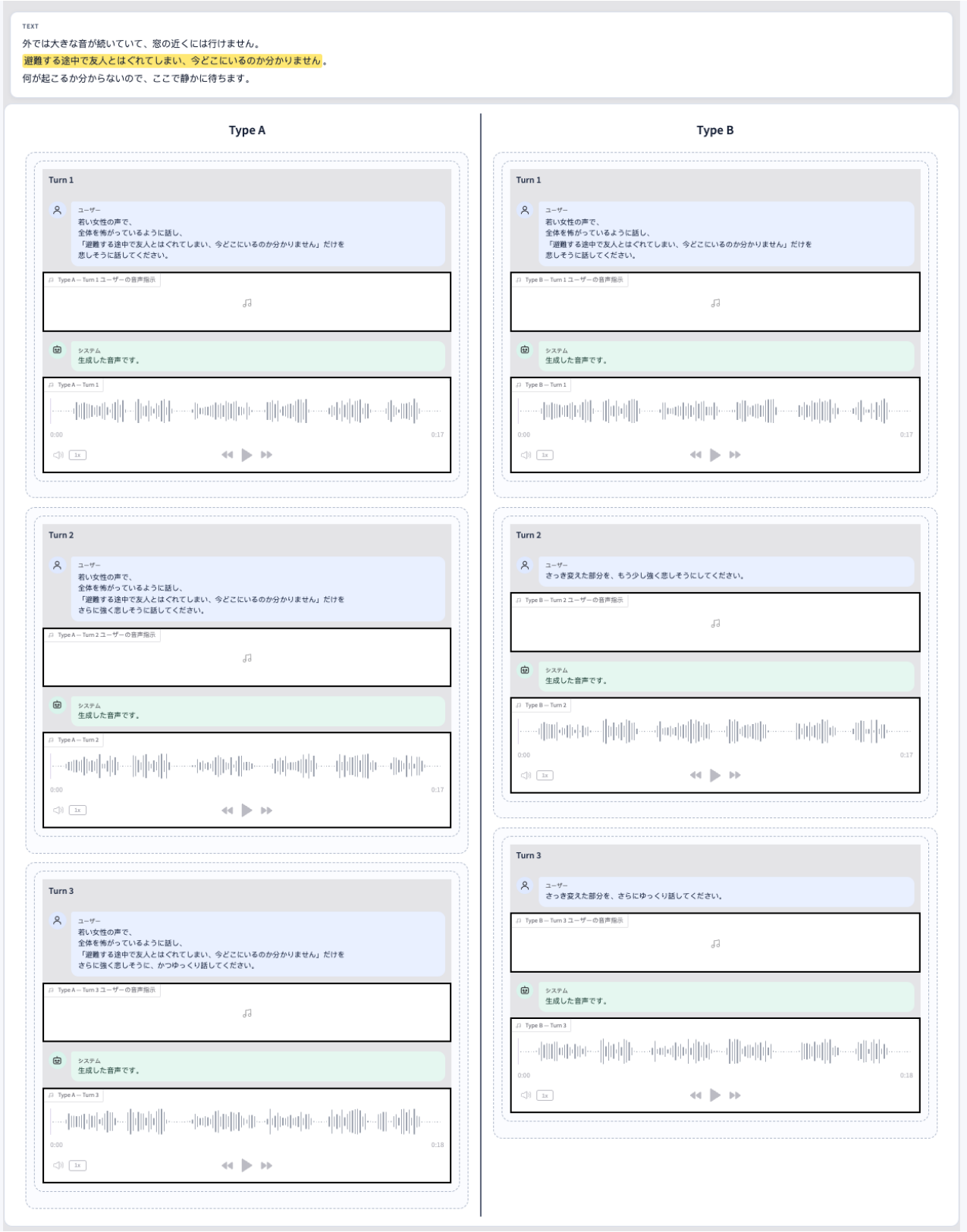}

    \caption{Interface used for the multi-turn evaluation.}
    \label{fig:multi}
\end{figure*}

Participants were native Japanese speakers of typical university-student age. Each evaluation session took approximately 90 minutes, and participants were compensated in accordance with the approved ethics policy. 
Figure \ref{fig:global}, Figure \ref{fig:tem}, and Figure \ref{fig:multi} show the interfaces used for the
global-style, temporal-control, and multi-turn evaluations. English translations of the interface instructions are on our demo page: \url{https://anonymous.4open.science/w/eacl-demo-17C5/}.

\paragraph{Global-style evaluation}
Each trial presented one source text, one speaking-style instruction,
and four anonymized speech samples. For example, the source text was
\jp{「ここだけの話ですが、明日の予定が少し変わりました。」}
(\textit{Koko dake no hanashi desu ga, ashita no yotei ga sukoshi
kawarimashita}; ``This is just between us, but tomorrow's schedule has
changed slightly.''). The instruction was
\jp{「若い女性が、小さな声で、ささやくように話してください。」}
(\textit{Wakai josei ga, chiisana koe de, sasayaku yō ni hanashite
kudasai}; ``Please speak quietly and whisper using a young female
voice.''). Listeners independently rated each sample for naturalness
(MOS) and agreement with the instruction (Match) on five-point scales.

\paragraph{Temporal-control evaluation}
Each trial highlighted the target span within the source text and
presented two anonymized samples. One example source text was
\jp{「ただ、問題が無事に解決したと聞いた瞬間は思わず笑顔に
なりました。それでも、担当者の無責任な対応には今でも納得して
おらず、きちんと説明を求めるつもりです。」}
(\textit{Tada, mondai ga buji ni kaiketsu shita to kiita shunkan wa
omowazu egao ni narimashita. Soredemo, tantōsha no musekinin na taiō
ni wa ima demo nattoku shite orazu, kichinto setsumei o motomeru
tsumori desu}; ``I could not help smiling when I heard that the problem
had been resolved. Even so, I remain dissatisfied with the staff
member's irresponsible response and intend to request a proper
explanation.'').

The instruction was
\jp{「若い女性話者が、全体を怒った口調で話す。さらに、
[TARGET: ただ、問題が無事に解決したと聞いた瞬間は思わず
笑顔になりました。]を非常に強い喜びの口調で話す。」}
(\textit{Wakai josei washa ga, zentai o okotta kuchō de hanasu.
Sarani, [TARGET: Tada, mondai ga buji ni kaiketsu shita to kiita
shunkan wa omowazu egao ni narimashita.] o hijō ni tsuyoi yorokobi no
kuchō de hanasu}; ``Use a young female voice and speak the entire
utterance angrily. Speak the target span with a very strong joyful
tone.'').

Listeners first selected the sample that better realized the target
style while preserving the non-target regions. They then rated only the
selected sample for overall naturalness, target-style accuracy,
non-target preservation, and transition naturalness on five-point
scales.

\paragraph{Multi-turn evaluation.}
Participants compared two three-turn revision sequences using the same
source text:
\jp{「外では大きな音が続いていて、窓の近くには行けません。
避難する途中で友人とはぐれてしまい、今どこにいるのか
わかりません。何が起こるかわからないので、ここで静かに
待ちます。」}
(\textit{Soto de wa ōkina oto ga tsuzuite ite, mado no chikaku ni wa
ikemasen. Hinan suru tochū de yūjin to hagurete shimai, ima doko ni iru
no ka wakarimasen. Nani ga okoru ka wakaranai node, koko de shizuka ni
machimasu}; ``Loud noises continue outside, so I cannot go near the
window. I became separated from my friend while evacuating and do not
know where they are. I do not know what will happen, so I will wait
quietly here.'').

The target span was
\jp{「避難する途中で友人とはぐれてしまい、今どこにいるのか
わかりません。」}
(\textit{Hinan suru tochū de yūjin to hagurete shimai, ima doko ni iru
no ka wakarimasen}; ``I became separated from my friend while
evacuating and do not know where they are.'').

Table~\ref{tab:multiturn_instruction_comparison} contrasts the complete
restatement and multi-turn instructions used at each turn.

\begin{table*}[t]
    \centering
    \caption{Instructions used in the full-restatement and multi-turn conditions.}
    \label{tab:multiturn_instruction_comparison}
    \footnotesize
    \setlength{\tabcolsep}{5pt}
    \renewcommand{\arraystretch}{1.12}

    \begin{tabularx}{\textwidth}{@{}
        c
        >{\raggedright\arraybackslash}X
        >{\raggedright\arraybackslash}X
        @{}}
        \toprule
        \textbf{Turn}
        & \textbf{Full Restatement}
        & \textbf{Multi-turn} \\
        \midrule

        \textbf{1}
        &
        \textbf{JP:}
        \jp{「若い女性の声で、全体を怖がっているように話し、
        『避難する途中で友人とはぐれてしまい、今どこにいるのか
        わかりません』だけを悲しそうに話してください。」}
        \newline
        \textbf{Romaji:}
        \textit{Wakai josei no koe de, zentai o kowagatte iru yō ni
        hanashi, `Hinan suru tochū de yūjin to hagurete shimai, ima doko
        ni iru no ka wakarimasen' dake o kanashisō ni hanashite kudasai.}
        \newline
        \textbf{EN:}
        Using a young female voice, speak the entire utterance fearfully
        and only the target span sadly.
        &
        \textbf{JP:}
        \jp{「若い女性の声で、全体を怖がっているように話し、
        『避難する途中で友人とはぐれてしまい、今どこにいるのか
        わかりません』だけを悲しそうに話してください。」}
        \newline
        \textbf{Romaji:}
        \textit{Wakai josei no koe de, zentai o kowagatte iru yō ni
        hanashi, `Hinan suru tochū de yūjin to hagurete shimai, ima doko
        ni iru no ka wakarimasen' dake o kanashisō ni hanashite kudasai.}
        \newline
        \textbf{EN:}
        Using a young female voice, speak the entire utterance fearfully
        and only the target span sadly.
        \\

        \midrule

        \textbf{2}
        &
        \textbf{JP:}
        \jp{「若い女性の声で、全体を怖がっているように話し、
        『避難する途中で友人とはぐれてしまい、今どこにいるのか
        わかりません』だけをさらに強く悲しそうに話してください。」}
        \newline
        \textbf{Romaji:}
        \textit{Wakai josei no koe de, zentai o kowagatte iru yō ni
        hanashi, `Hinan suru tochū de yūjin to hagurete shimai, ima doko
        ni iru no ka wakarimasen' dake o sara ni tsuyoku kanashisō ni
        hanashite kudasai.}
        \newline
        \textbf{EN:}
        Using a young female voice, speak the entire utterance fearfully
        and make only the target span sound even sadder.
        &
        \textbf{JP:}
        \jp{「さっき変えた部分を、もう少し強く悲しそうにしてください。」}
        \newline
        \textbf{Romaji:}
        \textit{Sakki kaeta bubun o, mō sukoshi tsuyoku kanashisō ni
        shite kudasai.}
        \newline
        \textbf{EN:}
        Make the previously modified part sound slightly sadder.
        \\

        \midrule

        \textbf{3}
        &
        \textbf{JP:}
        \jp{「若い女性の声で、全体を怖がっているように話し、
        『避難する途中で友人とはぐれてしまい、今どこにいるのか
        わかりません』だけをさらに強く悲しそうに、かつゆっくり
        話してください。」}
        \newline
        \textbf{Romaji:}
        \textit{Wakai josei no koe de, zentai o kowagatte iru yō ni
        hanashi, `Hinan suru tochū de yūjin to hagurete shimai, ima doko
        ni iru no ka wakarimasen' dake o sara ni tsuyoku kanashisō ni,
        katsu yukkuri hanashite kudasai.}
        \newline
        \textbf{EN:}
        Using a young female voice, speak the entire utterance fearfully
        and make only the target span sound even sadder and slower.
        &
        \textbf{JP:}
        \jp{「さっき変えた部分を、さらにゆっくり話してください。」}
        \newline
        \textbf{Romaji:}
        \textit{Sakki kaeta bubun o, sara ni yukkuri hanashite kudasai.}
        \newline
        \textbf{EN:}
        Speak the previously modified part even more slowly.
        \\

        \bottomrule
    \end{tabularx}
\end{table*}

The full-restatement condition explicitly repeated all accumulated
controls at every turn, whereas the incremental condition relied on
dialogue context and specified only the newly requested change.
Participants rated both sequences for perceived convenience,
interaction comfort, and modification retention on five-point scales.
After completing all dialogues, they selected full restatement,
incremental revision, or either condition as their overall preference.

\end{document}